\documentstyle[aps]{revtex}
\begin{document}
\title{Interacting hard-core bosons and surface preroughening}
\vspace{10mm}
\author{Alessandro Laio$^{1,3}$, Giuseppe Santoro$^{1,3}$, and
Erio Tosatti$^{1,2,3}$}
\address{
$^{(1)}$ International School for Advanced Studies (SISSA), Via Beirut 2,
Trieste, Italy\\
$^{(2)}$ International Center for Theoretical Physics (ICTP), Strada Costiera,
Trieste, Italy\\
$^{(3)}$ Istituto Nazionale per la Fisica della Materia (INFM), 
Via Beirut 2, Trieste, Italy
}
\maketitle
\begin{abstract}
The theory of the preroughening transition of an unreconstructed surface, 
and the ensuing disordered flat (DOF) phase, is formulated in terms of 
interacting steps. 
Finite terraces play a crucial role in the formulation.
We start by mapping the statistical mechanics of interacting (up and down) 
steps onto the quantum mechanics of two species of one-dimensional 
hard-core bosons.
The effect of finite terraces translates into a number-non-conserving term
in the boson Hamiltonian, which does not allow a description in terms of
fermions, but leads to a two-chain spin problem. 
The Heisenberg spin-1 chain is recovered as a special limiting case. 
The global phase diagram is rich. 
We find the DOF phase is stabilized by short-range repulsions of like steps. 
On-site repulsion of up-down steps is essential in producing a DOF phase, 
whereas an off-site attraction between them is favorable but not required. 
Step-step correlation functions and terrace width distributions can be directly 
calculated with this method.
\end{abstract}
%
%%%%%%%%%%%%%%%%%%%%%%%%%%%%%%%%%%%%%%%%%%%%%%%%%%%%%%%%%%%%%%%%%%%%%%%%%%
%                               TEXT
%%%%%%%%%%%%%%%%%%%%%%%%%%%%%%%%%%%%%%%%%%%%%%%%%%%%%%%%%%%%%%%%%%%%%%%%%%
\newpage
\section{Introduction} \label{intro:sec}

The surface roughening transition and the nature of the rough phase are
theoretically very well understood by a variety of approaches ranging from
phenomenological descriptions, based on the sine-Gordon model, to
microscopic solid-on-solid (SOS) models.\cite{Weeks}

The preroughening transition (PR) and the ensuing
disordered flat phase (DOF), both predicted several years ago by Rommelse
and den Nijs,\cite{Rom_MdN} have also been studied and characterized within
certain restricted solid-on-solid (RSOS) models. 
\cite{MdN_Rom,Jas,Mazzeo,Santoro,Goodstein,Prestip,Bastia} 
Although the physics behind these models, and the ingredients stabilizing the 
DOF phase, have been discussed in some detail, it is still useful to 
explore this subject from a different, and perhaps more physically appealing,
perspective. 
The RSOS models, in particular, do not directly emphasize steps, terraces, 
and kinks, which on the other hand are very crucial actors in these 
transitions.
This need is made more urgent by the recent experimental evidences for 
preroughening on rare gas solid (111) surfaces,\cite{Youn,Day}
which calls for a detailed reinvestigation of the step-step interactions, 
\cite{Prestip} or the reconstructive tendencies, \cite{Mazzeo,Santoro,Bastia}
crucial to obtain a DOF phase.

In the context of roughening, Villain and Vilfan,\cite{VV_88,VV_90,VV_91} 
den Nijs,\cite{MdN_92} and Balents and Kardar,\cite{Bal_Kar} for the case of
reconstructed surfaces, and Villain, Grempel, and Lapujoulade,\cite{VGL} 
for the case of vicinal surfaces, have shown that a more phenomenological 
approach based on working directly with steps yields a very direct picture 
of the physics involved. 

Anisotropic surfaces, in particular, have a definite direction of stronger 
bonding along which a step tends to run; kinks on such steps,
involving the breaking of strong bonds, are energetically expensive.
The (110) face of fcc noble metals (Au,Pt,Ag,$\dots$) is a physical
realization of an anisotropic surface, with a stronger, compact, 
$[1\bar{1}0]$ direction, and a softer $[001]$ direction. 
In the strongly anisotropic limit, the transfer matrix problem for the system 
of steps can be mapped onto the imaginary-time evolution of a system of quantum 
particles in one dimension (1D), the imaginary-time being the preferred 
direction in which the steps run.\cite{Kogut,Nijs_review,Schulz} 
This is a well known mapping, heavily exploited, for instance, 
in the theory of uniaxial commensurate-to-incommensurate transitions of 
adsorbates.\cite{Schulz,Nijs_review} 

Following a line of thought initiated by Ref.\ \cite{VV_90}, Balents and 
Kardar applied the full machinery of the theory of interacting fermions in one 
dimension to explore the possible phase diagrams of generic 
$(p\times 1)$ reconstructed anisotropic surfaces.\cite{Bal_Kar}
In their approach, steps of double height are forbidden (as 
energetically too expensive), while up and down monoatomic steps are mapped 
onto spin-1/2 fermions in 1D, described (in the continuum limit) 
by the hamiltonian \cite{Bal_Kar}
\begin{equation} \label{BK_1:eqn}
H \,=\, \sum_{\sigma} \int \! dx \psi_{\sigma}^{\dagger} 
(-\frac{\gamma}{2} \partial_x^2 -\mu ) \psi_{\sigma} 
+ \sum_{\sigma \sigma'} \int \! dx \, dx' n_{\sigma}(x) V_{\sigma \sigma'}(x-x')
n_{\sigma'}(x') \;.
\end{equation}
Here $\gamma$ is the inverse line tension of a step, $V_{\sigma \sigma'}(x)$
is an interaction between steps, and the remaining notation is standard.
This Hamiltonian describes infinite steps, traversing the entire length
of the sample, as the number of particles is conserved by the 
(imaginary-time) dynamics.
In reality, steps on surfaces can lead to finite defects by forming loops 
(i.e., finite terraces) on the surface.\cite{VV_90,Bal_Kar}
The order of the reconstruction $p$, dictates, through the symmetry of the
different ground states, the form of the ``loop'' terms which are 
allowed.\cite{VV_90,Bal_Kar}
For a $(p\times 1)$ reconstructed surface, the Hamiltonian $H$ has to be 
supplemented with a term of the type 
\begin{equation} \label{BK_2:eqn}
H_{\rm LOOP} = \lambda \int \! dx \left[ \prod_{k=0}^{p-1} 
\psi_{\uparrow}(x+ka) \psi_{\downarrow}(x+ka) \;+\; {\rm H.c.} \right] \;,
\end{equation}
where $a$ is a cut-off distance of the order of the lattice constant.
Balents and Kardar argued, by power counting, that the effect of finite 
terraces, i.e., the introduction of $H_{\rm LOOP}$, is irrelevant for $p>2$, 
marginal for $p=2$, and strongly relevant for $p=1$.
They went on by addressing in detail the $p=2$ case, of relevance to
the (110) missing-row reconstructed facet of Au. 
(See also Ref.\ \cite{MdN_92} for closely related work on the $p=2$ case.) 
However, the unreconstructed ($p=1$) case was not pursued further. 

The approach we take in the present paper is similar in spirit.
Our specific goal, however, is to address the question of the presence of
a DOF phase, and the preroughening transition, for unreconstructed surfaces. 
Thus, in the classification introduced above, we are now interested in detail
in the $p=1$ case. 
Technically, this leads, as we shall see, to significant differences 
with respect, for instance, to Ref.\ \cite{Bal_Kar},
and to a new phase diagram quite different from the $p>1$ cases. 

The crucial point is that for $p=1$ the loop terms (see Eq.\ (\ref{BK_2:eqn})) 
are of the BCS-like form 
$\lambda \int dx (\psi_{\uparrow}(x) \psi_{\downarrow}(x) + {\rm H.c.})$, 
i.e., a strongly relevant one-body piece.
This might appear as just a minor complication, at first glance, since quadratic
terms can be easily diagonalized by a Bogoliubov transformation. 
Closer consideration, however, leads to reconsider the whole mapping.
Fermionic minus signs have no role whatsoever in the mapping of a classical 
statistical mechanics problem. 
The natural statistics to use is always the bosonic. 
In the present case, a hard-core constraint will be necessary, in order to 
implement the appropriate configurational space 
(for instance, the non-crossing constraint for steps of the same type). 
In a one-dimensional quantum problem,
the choice of the statistics is, quite often, not a big problem, as we can 
transform, by a Wigner-Jordan transformation, hard-core bosons into fermions, 
with a transformed hamiltonian which has exactly the same form 
(only boundary conditions have to be considered carefully). 
In our case, however, pairing terms of the type 
$a_{i,\uparrow} a_{i+1,\downarrow}$, 
which (see below) are essential to describe finite terraces, 
do not transform into simple fermionic BCS-like terms, and become non-local 
after a Wigner-Jordan transformation. 
This will force us to work with hard-core bosons. 

Our approach, in summary, is as follows. 
We assume, as in Ref.\ \cite{Bal_Kar}, that the only relevant extended defects 
are monoatomic steps, which can be either {\it up} or {\it down}.
Steps of the same kind are forbidden to cross, while steps of 
different type can cross.
Moreover, steps interact with each other, have kinks, and can form finite 
terraces on the surface. 
These steps are then mapped onto world-lines of hard-core bosons in one 
dimension. 
Kinks on the steps correspond to hopping terms in the quantum Hamiltonian. 
Pairs of up-down steps which are created and annihilated to form finite 
terraces on the surface, give number-non-conserving terms in the quantum 
Hamiltonian.\cite{Schulz} 
Pairwise interactions between the steps are taken into
account by corresponding two-body terms in the quantum model.

The present work is concerned with the case of a low-index unreconstructed 
surface. Extensions to the case of vicinals, for which the long-ranged nature
of the step-step interactions is an essential ingredient, are left to a 
future study.

Our main goals in working out this type of approach to PR are the following:
(a) to build a formulation providing a more direct access to the physics of
PR, which is somewhat hidden in the RSOS formulations; 
(b) to explore more directly the role of step-step interactions; 
(c) to study step-step correlation functions and terrace 
width distributions, not available so far.
As it turns out, we have found that this approach is quite successful on all 
three accounts. 

The paper is organized as follows.
In Sec.\ \ref{model:sec} we present in detail the classical statistical 
mechanics model of interacting steps, which is then mapped onto the 
corresponding 1D quantum model of hard-core bosons
in Sec.\ \ref{quantum_model:sec}.
In Sec.\ \ref{canonical:sec} we consider in detail the spin-1 limit, 
obtained by setting the on-site step-step repulsion to infinity, 
and then map the general case to a problem of two spin-1/2 Heisenberg chains.
Sec.\ \ref{K:sec} contains a summary of bosonization plus
finite-size scaling calculations done in order to extract the phase diagram
of the model. 
Sec.\ \ref{order_parameters:sec} summarizes the relevant order 
parameters and correlation functions investigated.  
In Sec.\ \ref{phase_diagram:sec} we present our results for the overall phase
diagram of the model.
Sec.\ \ref{corr:sec} illustrates our results for the step-step correlations
and the terrace width distributions.
Finally, Sec.\ \ref{conclusions:sec} contains a discussion of the results
and some conclusive remarks. 

%+++++++++++++++++++++++++++++++++++++++++++++++++++++++++++++++++++
\section{The model} \label{model:sec}
%+++++++++++++++++++++++++++++++++++++++++++++++++++++++++++++++++++

We assume the only relevant extended defects involved in the surface PR
transition to be steps, which can be either {\it up} or {\it down}. These
steps interact with each other, they can have kinks, and they can form 
finite terraces on the surface. 
Fig.\ \ref{step:fig} shows a schematic picture of a surface with steps.

Our model will be defined on a square lattice, and we will assume the steps
to run preferentially in one direction 
(the vertical direction in Fig.\ \ref {step:fig}). 
The steps are only allowed to make simple nearest neighbor kinks.
Steps running in the horizontal direction are assumed to be
energetically expensive and neglected. Hence, our surface is, by construction, 
very highly anisotropic. 

We define the model by assigning its transfer matrix along y.
Denoting by $\left| \rm{S}(j) \right\rangle$ and 
$\left| \rm{S}(j+1) \right\rangle$ the configuration of the $j$-th and
$j+1$-th horizontal strips, we have 
\[
\left\langle \rm{S}\left( j+1\right) \right| \cal{T}\; \left| 
\rm{S}\left( j\right) \right\rangle = 
\]
\begin{equation}
= \exp \left( -\beta \left( 
\delta_S N_S^{(j)} + \delta_K N_K^{(j,j+1)}
+\sum_{s=0,1} \delta_{T_s} N_{T_s}^{(j,j+1)}
+\delta_{ex} N_{ex}^{(j,j+1)}
+V_{\rm step-step} \right) \right)  
\label{Tmatrix}
\end{equation}
where $\beta$ is the inverse temperature, and 
(see also Fig.\ \ref{transfer:fig}):

\begin{itemize}
\item  $\delta_S$ represents the energy cost (per unit length) of a step
running along the y-direction; $N_S^{(j)}$ is the number of
steps in the strip $j$;

\item  $\delta_K N_K^{(j,j+1)}$ is the energy cost of 
$N_K^{(j,j+1)}$ kinks between strip $j$ and $j+1$;

\item  $\delta_{T_s} N_{T_s}^{(j,j+1)}$ is the energy cost for the creation 
of $N_{T_s}^{(j,j+1)}$ terraces of ``size'' $s$ between strip $j$ and $j+1$.
We will always assume $s=1$, or $0$ (see Fig.\ \ref{transfer:fig}). 
 
\item  $\delta_{ex} N_{ex}^{(j,j+1)}$ is the energy associated to the 
crossing of $N_{ex}^{(j,j+1)}$ pairs of opposite steps between strip $j$ and 
$j+1$;

\item  $V_{\rm step-step}=V^{\parallel}+V^{\perp}$, with 
$V^{\parallel}$ and $V^{\perp}$ describing respectively the interaction
between steps of the same kind and of the opposite kind. 
For $V^{\parallel}$, we assume a generic repulsive interaction 
\begin{equation}
V^{\parallel} = \sum_{\sigma} \sum_{k>i}\widetilde{V}_{k-i}^{\parallel}
n_{i,\sigma} n_{k,\sigma} \;,  
\label{vparallel}
\end{equation}
with $V_{k-i}^{\parallel}$ possibly possessing an elastic long-range tail 
of the form $\approx |k-i|^{-2}$. Here $n_{i,\uparrow (\downarrow )}$
is $1$ if there is a step up (down) at site $i$. 
Similarly, we assume $V^{\perp}$ to be given by 
\begin{equation}
V^{\perp} = \widetilde{V}_0^{\perp} \sum_i n_{i,\uparrow} n_{i,\downarrow}
+ \sum_{\sigma} \sum_{k>i} \widetilde{V}_{k-i}^{\perp } n_{i,\sigma}
n_{k,\bar{\sigma}} \;.  
\label{vperp}
\end{equation}
The sign of the terms in $V^{\perp}$, particularly at short-range, depends
on microscopic details and need not be specified at this stage.
\end{itemize}

If we assume periodic boundary conditions in $y$ direction, i.e., 
$\left| \rm{S}(N_y+1) \right\rangle = \left| \rm{S}(1) \right\rangle$, 
the partition function of this system is
\begin{equation}
Z \,=\, \lim_{N_y\to \infty }\,\rm{Tr}\;\;{\cal{T}}^{N_y}\;.
\end{equation}

%+++++++++++++++++++++++++++++++++++++++++++++++++++++++++++++++++++
\section{The quantum model} \label{quantum_model:sec}
%+++++++++++++++++++++++++++++++++++++++++++++++++++++++++++++++++++

It is well known that, in the strong anisotropy (or time-continuum) limit,
many $D$ dimensional classical problems can be mapped onto $D-1$ dimensional
quantum problems.\cite{Kogut} 
This relationship is established by means of the path
integral formalism. In particular, the up and down steps are mathematically
equivalent to the world lines of spin-up and spin-down hard-core bosons, and
the preferential direction in which the steps run plays the role of time
in the quantum problem. The hard-core condition is imposed in order to 
implement the non-crossing condition for steps of the same type, 
a physically justified restriction, in view of the large energetic cost of 
double-step regions.
The non-crossing constraint for steps of the same type would
be automatically satisfied by the Pauli principle if we were to deal with
spin-1/2 fermions. 
(See below for more comments on the problem of quantum statistics.)

We consider the following quantum Hamiltonian 
\[
H = -t^{\parallel} \sum_{i,\sigma} (a_{i,\sigma}^{\dagger} a_{i+1,\sigma}
+ {\rm{H.c.}}) -\mu \hat{N} \,+ 
\]
\[
- t_0^{\perp} \sum_i (a_{i,\uparrow}^{\dagger} a_{i,\downarrow}^{\dagger}
+ {\rm{H.c.}}) 
- t_1^{\perp} \sum_{i,\sigma} 
   (a_{i,\sigma}^{\dagger} a_{i+1,\bar{\sigma}}^{\dagger} + {\rm{H.c.}}) \,
- t_{ex} \sum_{i,\sigma} 
  a_{i+1,\overline{\sigma}}^{\dagger} a_{i,\sigma}^{\dagger}
  a_{i+1,\sigma} a_{i,\overline{\sigma}} \,+ 
\]
\begin{equation}
+ \sum_\sigma \sum_{j>i} V_{j-i}^{\parallel} n_{i,\sigma} n_{j,\sigma}
+ V_0^{\perp} \sum_i n_{i\uparrow} n_{i\downarrow} 
+\sum_\sigma \sum_{j>i} V_{j-i}^{\perp} n_{i,\sigma } n_{j,\bar{\sigma}}
\label{hamiltonian}
\end{equation}
with $a_{i,\sigma}$ representing the destruction operator for a spin 
$\sigma $ hard-core boson and with 
$\hat{N}=\hat{N}_{\uparrow }+\hat{N}_{\downarrow}$ the total number 
of particles. 
We will work in the subspace $\hat{N}_{\uparrow}=\hat{N}_{\downarrow}$,
for a low-index surface.
(For a vicinal surface of angle $\phi$, we would have 
$(\hat{N}_{\downarrow}-\hat{N}_{\uparrow})= L\tan{\phi}$.)

Within a path-integral approach, it can be shown \cite{Kogut}
that the {\it ground state} properties of this quantum Hamiltonian correspond
to the {\it temperature} properties of the classical step model, whose
transfer matrix is given by Eq.\ (\ref{Tmatrix}), 
in the large anisotropy limit. 
Specifically, the classical parameters turn out to be given by:
\begin{eqnarray}
\epsilon \,t^{\parallel } &=& e^{-\beta \delta_K},\hspace{5mm}
\epsilon \,t_{0,1}^{\perp }=e^{-\beta \delta_{T_{0,1}}},\hspace{5mm}
\epsilon \,t_{ex}=e^{-\beta \delta _{ex}}  \nonumber  \\
\epsilon (-\mu ) &=& \beta \delta_S,\hspace{5mm} \epsilon 
V_{i-j}^{(\parallel,\perp)}=\beta \tilde{V}_{i-j}^{(\parallel,\perp)}  
\label{parameters}
\end{eqnarray}
where $\epsilon$ is the Trotter discretization time for the quantum 
path-integral. 

The mapping is asymptotically correct only in the limit $\epsilon \to 0$. 
This a) implies, clearly, a {\it strong anisotropy limit\/} for the 
classical problem and b) does not allow a straightforward
identification of a classical low-- or high--temperature limit. 
Indeed, if all the parameters of the quantum problem are of 
order one, taking $\beta \to 0$ or $\beta \to \infty$ makes 
Eq.\ (\ref{parameters}) incompatible with the requirement that the left-hand 
side should be a small quantity (of order $\epsilon$). 
In other words, the mapping is justified so long as 
\[
\mbox{Kinetic couplings}=(\delta _K,\delta _{T_0},\cdots )\gg T\gg (\delta
_S,\tilde{V},\cdots )=\mbox{Potential couplings}\;, 
\]
and nothing can be said, in principle, about the infinite temperature limit. 
This should be always kept in mind when considering the infinite
temperature limit from the quantum model point of view 
(see, e.g., sec.\ \ref{XY:sec}). 

It is also worth stressing that the (hard-core) boson statistics of the 
$a$ operators in the Hamiltonian is crucial to the nature of the phases 
and transition lines in the phase diagram. 
Indeed, unlike other terms in the Hamiltonian, the terrace creation
terms cannot be translated into simple (i.e., local) fermionic BCS-like
terms by a Jordan-Wigner transformation. 
In such instances, the correct statistics to use is undoubtedly the bosonic 
one, as fermionic minus signs do not appear in a classical statistical 
mechanics problem.
This point seems to be not always appreciated in the literature.\cite{Kar_Sha}

We now make contact with previous work in the context of surface physics. 
The model in Eq.\ (\ref{hamiltonian}), with $t_{ex}=0$ and 
$t_{0,1}^{\perp}=0$, has been considered, in its continuum version, 
by Balents and Kardar.\cite{Bal_Kar} 
(See also Refs.\ \cite{VV_88} and \cite{MdN_92} for related work.)
In the absence of $t^{\perp}$-terms, particles are taken to be fermions. 
The emphasis of Ref.\ \cite{Bal_Kar} was on $(p\times 1)$
reconstructed surfaces, particularly with $p=2$. 
The effect of finite terraces, i.e., closed loops of steps, was argued to be 
irrelevant for $p>2$, marginal for $p=2$, and strongly relevant for 
$p=1$.\cite{Bal_Kar}
The unreconstructed ($p=1$) case, however, was not pursued at all. 
As just argued, the $p=1$ case cannot be tackled in terms of fermions. 
The effect of the finite terraces on an unreconstructed surface 
-- the $t^{\perp}$ terms in Eq.\ (\ref{hamiltonian}) -- 
is one of the points addressed in detail in the present work. 
Moreover, we show that restricting the analysis to a simple Hubbard-type 
on-site interaction does not lead to the full richness of the phase diagram;
nearest-neighbor interactions are essential in order to stabilize, for 
instance, a DOF phase.

%+++++++++++++++++++++++++++++++++++++++++++++++++++++++++++++++++++
\section{Canonical Transformations} \label{canonical:sec}
%+++++++++++++++++++++++++++++++++++++++++++++++++++++++++++++++++++

%+++++++++++++++++++++++++++++++++++++++++++++++++++++++++++++++++++
\subsection{$V_0^{\perp}=\infty$: Mapping to a spin-1 chain.} \label{Spin-1:sec}
%+++++++++++++++++++++++++++++++++++++++++++++++++++++++++++++++++++

For a special choice of the parameters, the model in Eq.\ (\ref{hamiltonian}) 
reduces to a well studied problem. 
Consider the case in which both $V^{\parallel}$ and $V^{\perp}$ are 
truncated to nearest neighbors, 
$V_{j-i}^{\parallel,\perp}=V_1^{\parallel,\perp}\delta_{j,i+1}$, 
in the limit of infinite on-site repulsion of opposite steps, 
$V_0^{\perp}\rightarrow \infty$. 
The limit $V_0^{\perp}\rightarrow \infty$ enforces, in absence of $t_{ex}$, 
a non-crossing condition for opposite steps as well,
and allows only three states per site, which we can easily map onto a 
{\it spin-1} variable as follows: 
\begin{eqnarray}
\mbox{step up} \hspace{5mm} a_{i,\uparrow}^{\dagger} |0\rangle &=&
| S_i^z=+1\rangle \nonumber \\ 
\mbox{no step} \hspace{5mm} \hspace{3mm} |0\rangle &=& 
| S_i^z=0\rangle \nonumber \\ 
\mbox{step down} \hspace{5mm} a_{i,\downarrow }^{\dagger }|0\rangle &=&
| S_i^z=-1\rangle \;.
\end{eqnarray}
It is then straightforward to show that all possible matrix elements of $H$
in Eq.\ (\ref{hamiltonian}) coincide exactly with those of the spin-1
Heisenberg chain 
\begin{equation}
H_{\rm{Heis}}=-\frac{J_{xy}}2\sum_i (S_i^{+}S_{i+1}^{-}+{\rm H.c.})
+ J_z\sum_i S_i^z S_{i+1}^z + D\sum_i (S_i^z)^2\;,  
\label{H_Heis}
\end{equation}
if the following parameter choice is made: 
\begin{equation}  \label{Heiscond}
t^{\parallel }=t_1^{\perp }=J_{xy}\;, \hspace{4mm} t_{ex}=0\;,\hspace{4mm}
V_1^{\parallel }=-V_1^{\perp }=J_z \;,\hspace{4mm}-\mu =D \;.
\end{equation}
A nonzero $t_{ex}$ term, when present, translates into a quartic spin term 
\begin{equation}  \label{tex:eqn}
-t_{ex}\sum_i [(S_i^{+})^2(S_{i+1}^{-})^2+\rm{H.c.}] \;, 
\end{equation}
and will be shown to be relevant in stabilizing the rough phase for finite 
repulsive $V_1^{\parallel}$. 
The phase diagram for this special case, in the surface
physics relevant region $-\mu =D>0$, can be directly borrowed from the 
literature.\cite{MdN_Rom,giapponesi}

The Heisenberg spin-1 chain was also obtained by another route, namely by
the quantum mapping of RSOS models, 
by den Nijs and Rommelse, who gave a very detailed discussion of the surface
physics interpretation of the different phases.\cite{MdN_Rom} The region in
which a DOF phase is stabilized is found for $J_z>0$, i.e., it corresponds
to repulsive $V_1^{\parallel }$ and attractive $V_1^{\perp }$ 
(see Eq. (\ref{Heiscond})). 
The latter condition is, however, not crucial, as we shall see later on. 
The flatness of the DOF phase is directly related to the Haldane
gap in the spin-1 chain.\cite{MdN_Rom} 
If the interactions are much bigger than the cost
of a unit of step ($J_z\gg D$, see line (a) in Fig.\ \ref{h_ph:fig}), 
at low temperatures (i.e., for large $J_z/J_{xy}$) the system will reconstruct 
into an ordered sequence of up-down steps (the N\'{e}el phase of the spin-1 
chain). 
By increasing temperature (i.e., lowering $J_z/J_{xy}$) the system undergoes 
an Ising transition to a DOF phase, in which the positional order of the 
up-down sequence of steps is lost.\cite{MdN_Rom} 
If the cost of the unit step is larger than the interactions, 
($J_z\ll D$, see line (b) in Fig.\ \ref{h_ph:fig}), the low temperature phase 
is flat, and the (preroughening) transition to the DOF phase has non-universal 
exponents.\cite{MdN_Rom}

If we impose the condition $V_0^{\perp}\rightarrow \infty$ without
assuming Eq.\ (\ref{Heiscond}), what we are considering is always a 
three-state-per-site problem, but the resulting Hamiltonian does not have the
simple bilinear form (\ref{H_Heis}) in terms of the spin-1 operators. 
We will refer to this general case as a {\it spin-1 chain}. 
The specific case in Eq.\ (\ref{H_Heis}) will be referred to as 
{\it Heisenberg spin-1 chain}.

%+++++++++++++++++++++++++++++++++++++++++++++++++++++++++++++++++++
\subsection{The general case: Mapping to two spin-1/2 chains} 
\label{2Spin_1/2:sec}
%+++++++++++++++++++++++++++++++++++++++++++++++++++++++++++++++++++

In order to study the model in Eq.\ (\ref{hamiltonian}) for more general
parameter values, in particular for $V_0^{\perp }\neq \infty$, it is
convenient to abandon the spin-1 representation and map onto a problem of two
coupled spin-1/2 chains, where the total number of states per rung is four,
instead of three. Introducing the usual spin-1/2 representation for
each of the two species of hard-core bosons, 
\[
a_{i,\uparrow (\downarrow )}^{\dagger }=S_{i,1(2)}^{+}\;,\hspace{10mm}%
n_{i,\uparrow (\downarrow )}=S_{i,1(2)}^z+\frac 12\;, 
\]
and performing a $\pi$-rotation for the ${\bf S}_{i,2}$ spins around the
x-axis, which amounts to a particle-hole transformation for the down-bosons, 
$S_{i,2}^{\pm}\to S_{i,2}^{\mp}$ and $S_{i,2}^z\to -S_{i,2}^z$, one can
rewrite the Hamiltonian (\ref{hamiltonian}) as the following model of two
spin-1/2 chains ($\alpha=1,2$ denoting the chain) with opposite magnetic 
fields: 
\begin{eqnarray}  \label{spinhamiltonian}
H \,=\, && -t^{\parallel}\sum_{i,\alpha}
      (S_{i,\alpha }^{+}S_{i+1,\alpha }^{-}+{\rm{H.c.}})
+V_1^{\parallel}\sum_{i,\alpha} S_{i,\alpha}^z S_{i+1,\alpha}^z
+h\sum_i(S_{i,1}^z-S_{i,2}^z)  \nonumber \\
&&-t_0^{\perp}\sum_i (S_{i,1}^{+}S_{i,2}^{-}+{\rm{H.c.}})
-V_0^{\perp}\sum_i S_{i,1}^zS_{i,2}^z \,+  \nonumber \\
&&-t_1^{\perp}\sum_{i,\alpha} 
      (S_{i,\alpha}^{+}S_{i+1,\bar{\alpha}}^{-}+{\rm{H.c.}})
-V_1^{\perp}\sum_{i,\alpha} S_{i,\alpha}^z S_{i+1,\bar{\alpha}}^z 
\,+  \nonumber \\
&&-t_{ex}\sum_i \left( S_{i+1,1}^{+}S_{i,1}^{-}\,S_{i+1,2}^{+}S_{i,2}^{-}+
{\rm{H.c.}}\right) \;,
\end{eqnarray}
where, for simplicity, we have considered only interaction terms up to
nearest neighbors and we have omitted the constant 
$\frac 14[V^{\perp}(q=0) +V^{\parallel}(q=0)]$ 
(with $V^{\parallel (\perp)}(q)$ we denote the Fourier transforms of the 
potentials, and 
$V^{\parallel (\perp)}(q=0)=\sum_{j}{V}_{j-i}^{\parallel (\perp)}$). 
The magnetic field $h$ is related to the chemical potential $\mu$ in the
following way:
\[
h=-\mu +\frac 12\left[ V^{\perp }\left( q=0\right) +V^{\parallel }\left(
q=0\right) \right] =-\mu +(V_0^{\perp }/2+V_1^{\perp }+V_1^{\parallel }) \;.
\]
Notice that after the spin-rotation, performed to get a standard 
$S_1^{+}S_2^{-}$ coupling between the chains starting from the 
$a_{\uparrow}^{\dagger}a_{\downarrow}^{\dagger}$ boson term, the signs of the 
$V^{\perp}$ terms are all changed, since $S_{i,2}^z\to -S_{i,2}^z$. 
For the same reason, the chemical potential terms transform into opposite
magnetic fields for the two chains. 
The condition $\hat{N}_{\uparrow}=\hat{N}_{\downarrow}$
reads, after the canonical transformation, as zero total magnetization for
the spins, $\sum_i\left( S_{i,1}^z+S_{i,2}^z\right) =0$. \cite{vicinal:note}

%+++++++++++++++++++++++++++++++++++++++++++++++++++++++++++++++++++++++
\section{Low-energy Hamiltonian from finite-size diagonalization} \label{K:sec}
%+++++++++++++++++++++++++++++++++++++++++++++++++++++++++++++++++++++++

With a well defined quantum spin chain problem at hand, it is standard
practice to study its phase diagram by a combination of field-theoretical 
arguments and finite-size exact diagonalization data. 

At weak coupling, a standard field-theory approach to one dimensional 
quantum systems consists in applying bosonization techniques. 
This was done by Strong and Millis for the case of two coupled spin-1/2
Heisenberg chains, i.e., a special case of Eq.\ (\ref{spinhamiltonian}) 
with $t_1^{\perp}=V_1^{\perp}=0$, $t_{ex}=0$, $h=0$, and 
$V_0^{\perp}=2t_0^{\perp}=-J_K$.\cite{Strong} 
The procedure can be easily extended to our case. 
Introducing symmetric and antisymmetric combinations of the bosonic phase
fields, $\Theta_{N,_A^S}$ and $\Theta_{J,_A^S}$, \cite{Strong,Haldane}
which represent the bosonic sound-like excitations of the system in the 
gapless phase, the low-energy Hamiltonian reads:
\[
H \,=\, \sum_{\alpha =S,A}\frac{v_\alpha }{4\pi} \int dx\,
\left[ \frac{1}{K_\alpha} \left( \nabla \Theta_{J,\alpha} \right)^2
+ K_{\alpha} \left( \nabla \Theta_{N,\alpha} \right)^2 \right] \,+ 
\]
\[
-\frac{1}{(\pi a)^2} \int dx\, 
\left\{ \cos \left( \Theta_{N,A} \right) 
     \left\{ A + B \left[ \cos \left( 2\Theta_{J,S} \right) 
                         +\cos \left( 2\Theta_{J,A} \right)
                   \right] 
\right\} \,+ 
C \cos \left( 2\Theta_{J,S} \right) + D \cos \left( 2\Theta_{J,A} \right) 
\right\} \,, 
\]
where $a$ is a short distance cut-off, and $A$, $B$, $C$, and $D$ are
coupling dependent (but cut-off independent) constants.

Let us consider the A-sector first. The A-sector can be gapless only if 
$1/\sqrt{K_A}>2$ and, simultaneously, $2\sqrt{K_A}>2$. 
This is, of course, impossible. 
Thus, the A-sector flows to strong coupling and develops a gap.\cite{Strong}

The Hamiltonian for the S-sector, renormalized by the A-sector, will be of
the form: 
\[
H_S \,=\, \frac{v_S}{4\pi} \int dx\, 
\left[ \frac{1}{K_S}\left( \nabla \Theta_{J,S} \right)^2 \,+\,
                K_S \left( \nabla \Theta_{N,S} \right)^2
\right] \,-\,
\frac{V_S}{(\pi a)^2} \int dx\, \cos \left( 2\Theta_{J,S} \right) \;. 
\]
The cosine term is relevant and opens up a gap when $K_S<1$. 
Thus the system undergoes a KT transition when $K_S\to 1$. 
As discussed in the following sections, this is associated to a 
roughening transition.

If the symmetric sector flows to the free Klein-Gordon Hamiltonian
(i.e., the Luttinger model) in some range of parameters, 
the low energy spectrum of the two chains, for $L\rightarrow \infty$, 
will have the form of a spinless Luttinger model describing symmetric
excitations. 
Expressing $\Theta_{J(N),S}$ in terms of bosonic creation operators 
we can write:\cite{Haldane}
\begin{equation}
H_S \,=\, v_S \sum_k\left| q\right| b_k^{\dag }b_k
\,+\, \frac{\pi}{4L} \left( v_N N_S^2 + v_J J_S^2 \right)  \;,
\label{HLutt2}
\end{equation}
where $N_S$ and $J_S$ are the symmetric sector total number ($N$) and 
current ($J$) operators, $v_S$ is the renormalized sound velocity, 
$v_N=v_S/K_S$, and $v_J=K_S v_S$.\cite{Haldane}

In order to compute $v_N$, we note that the simplest charge excitation not
involving the current part consists in adding $2$ particles to the system.
Thus we have: 
\[
v_N(L) = \frac L\pi \left[ E(\Delta N=2) -E(0) \right]  \;.
\]
To compute $v_J$, notice that if a magnetic flux $\Phi$ is
concatenated with the ring, the current part of the energy spectrum is
modified in this way: 
\[
\frac{\pi}{4L} v_J J_S^2 \longrightarrow 
\frac{\pi}{4L} v_J \left( J_S + 4\frac{\Phi}{\Phi_0} \right)^2 
\]
($\Phi _0$ is the elementary flux quantum). Therefore:
\[
v_J(L) = \frac{L}{8\pi} 
\frac{\partial^2 E(\Phi )}{\partial \left(\Phi /\Phi_0\right)^2 } \;. 
\]
Finally, $v_S$ can be computed from
\[
v_S(L) = \frac{L}{2\pi} \left[ E(k=2\pi/L) - E(0) \right] 
\]
where $E(0)$ is the ground state energy and $E(k=2\pi/L)$ is the energy of 
the lowest excited state of momentum $k=2\pi/L$.

As a consequence, $K_S$ can be equivalently computed from the finite-size
extrapolation of $v_J(L)/v_S(L)$, of $v_S(L)/v_N(L)$, or of 
$\sqrt{v_J(L)/v_N(L)}$. 
If the finite-size data are compatible with a Luttinger Liquid picture, 
i.e., with a spectrum of the form (\ref{HLutt2}), then these three 
extrapolations should converge, as $L\to \infty$, to a single value.

%+++++++++++++++++++++++++++++++++++++++++++++++++++++++++++++++++++
\section{Order parameters} \label{order_parameters:sec}
%+++++++++++++++++++++++++++++++++++++++++++++++++++++++++++++++++++

We now define the order parameters and correlation functions we have to
consider in order to study the phase diagram of our model. We will basically
deal with four correlation functions, whose behavior in the different phases
is summarized in Tab.\ 1.

\begin{itemize}
\item  The {\it height-height correlation}, defined by 
$G_h(r) = \langle ( h_r-h_0)^2 \rangle$, diverges logarithmically as 
\[
G_h\left( r\right) =\frac{2K}{\pi ^2}\ln \left( r\right) +\dots 
\]
in the rough phase, with $K\geq 1$, while it remains limited in the flat, 
N\'eel, and DOF phases. \cite{G_h:note}
At the roughening transition, $K$ takes the universal
value of $1$ \cite{Weeks,MdN_92}. This gives a simple criterion for
determining whether a phase is rough or not. 
In fact, the coefficient $K$ coincides with the Luttinger exponent $K_S$ 
for the symmetric sector,\cite{Bal_Kar} which can be extracted by finite-size 
scaling of exact diagonalization data (cfr.\ Sec.\ \ref{K:sec}). 

\item  The {\it string correlation function}, defined by \cite{MdN_Rom}
\[
G_s\left( r\right)=-\langle (S_0^z\exp{(i\pi\sum_{j=1}^{r-1}S_j^z)}S_r^z
\rangle \;.
\]
(We introduced the notation ${S_i^z=n}_{i\uparrow }-n_{i\downarrow }$.) 
The phase factor contributes a plus (minus) sign if there are an even (odd)
number of steps between site $0$ and site $r$. In the DOF and in the
N\'{e}el phases, a step up (down) is preferentially followed by a step
down (up). 
In these configurations, $G_s(r)$ gets a contribution equal to $1$
every time sites $0$ and $r$ are occupied by a step.
Thus, in the DOF phase and in the N\'{e}el phase $G_s(r)$ 
decays exponentially to the square of the mean density of steps.\cite{MdN_92}  

\item  The {\it staggered magnetization}, defined by
\[
{\cal N} =\lim_{L\rightarrow \infty }\frac 1L\sum_j(-1)^j
\langle S_0^zS_j^z\rangle \;. 
\]
A N\'{e}el phase will be signaled by a non-zero staggered magnetization 
$\cal{N}$, while $\cal{N}$ is zero in the rough, DOF and flat phases.

\item  The {\it flatness order parameter}, defined by \cite{MdN_Rom}
\[
{\cal F} =\lim_{L\rightarrow \infty } \frac 1L \sum_r
\langle \exp {(i\pi \sum_{j=1}^rS_j^z)}\rangle \;. 
\]
$\cal{F}$ has a non zero value only in the ordered flat phase (in the DOF phase 
the exponential fluctuates between $1$ and $-1$ as $r$ is increased).
\end{itemize}

\[
\begin{tabular}{|l|l|l|l|l|}
\hline
\thinspace  & \bf{Flat} & \bf{Rough} & \bf{DOF} & \bf{N\'eel} \\ \hline
$K_S$ & gapped & $>1$ & gapped & gapped \\ \hline
$G_h\left( r\right) \;\;\left( {\mbox{for large }}r\right) $ & $<\infty $ & $%
\rightarrow \frac{2K_S}{\pi ^2}\ln \left( r\right) $ & $<\infty $ & $<\infty 
$ \\ \hline
$G_s\left( r\right) \;\;\left( {\mbox{for large }}r\right) $ & $\rightarrow 0$
& $\rightarrow 0$ & $\rightarrow \langle \left( S^z\right) ^2\rangle $ & $%
\rightarrow \langle \left( S^z\right) ^2\rangle $ \\ \hline
$\cal{N}$ & $=0$ & $=0$ & $=0$ & $\neq 0$ \\ \hline
$\cal{F}$ & $>0$ & $0$ & $0$ & $0$ \\ \hline
\end{tabular}
\]

%+++++++++++++++++++++++++++++++++++++++++++++++++++++++++++++++++++
\section{The Overall Phase Diagram} \label{phase_diagram:sec}
%+++++++++++++++++++++++++++++++++++++++++++++++++++++++++++++++++++

Our model, even if the interactions are truncated to first neighbors,
contains many parameters. Rather than trying to describe the
phase diagram in an exhaustive form, we will focus our discussion on a few
questions, that we consider quite relevant with respect to the 
{\it surface physics} interpretation of our model. 
This will take us to consider in detail some special planes cut 
through the phase diagram, and will also give us an idea of 
its global structure.

Let us consider, once again, the Heisenberg spin-1 phase diagram in 
Fig.\ \ref{h_ph:fig}. 
From the surface physics point of view, it presents some
unpleasant features: since the increasing temperature curve, for a given 
surface, is a line through the origin (the origin corresponds to the infinite 
$T$ point; cfr.\ Eq.\ (\ref{parameters})), every ``surface'' with repulsive
interaction between steps of the same kind ($V_1^{\parallel}>0$)
has a preroughening transition at finite temperature, and no rough phase 
at finite $T$. On the other hand, if $V_1^{\parallel}$ is attractive
there is only roughening.

In relation to these problems, we will discuss the following main questions:

\begin{description}
\item[A] Is the attractive $V_1^{\perp}$ term between opposite sign
steps essential in order to stabilize a DOF phase?

\item[B] Is there a choice of the parameters for which our model can describe
a surface with a finite roughening temperature?

\item[C] How does the presence or absence of a preroughening transition 
depend on the relative strength of the step-step interactions to the cost 
per unit length of a step. 

\item[D] What is the role of the opposite step on-site repulsion $V_0^{\perp}$.

\end{description}

In the following we address these points directly.

%++++++++++++++++++++++++++++++++++++++++++++++++++++++++++++++++++++++
\subsection{The role of attraction between opposite steps: Spin-1 chain 
with $V_1^{\perp }=0$.} \label{V1_perp_0:sec}
%++++++++++++++++++++++++++++++++++++++++++++++++++++++++++++++++++++++

In order to explore the different roles of the two interactions $V_1^{\perp}$ 
and $V_1^{\parallel}$, we have first studied the effect of a repulsive 
$V_1^{\parallel}$, keeping $V_1^{\perp}=0$. 
For the time being, we still work with the spin-1 condition, i.e., we impose an
infinite on-site repulsion of opposite steps ($V_0^{\perp}=\infty$).

For this choice of parameters, we do not find any point with 
$V_1^{\parallel}>0$ in which the finite-size data might indicate a 
vanishing gap. 
This is compatible with the results of den Nijs and Rommelse about the 
location of the KT-transition in the Heisenberg spin-1 phase 
diagram.\cite{MdN_Rom}
The system undergoes a roughening transition only at infinite temperature. 

In Fig.\ \ref{pd_vpar:fig} we draw a qualitative phase diagram for values of
the parameters $\mu$ and $V^{\parallel}$ which are relevant for surface
physics (i.e., positive energy cost for a step, and repulsive interaction
between steps of the same kind). 
It is quite remarkable how the DOF phase survives the turning off of the
attraction between steps of the same kind. 
As a matter of fact, taking $V_1^{\perp}=0$ leads to the disappearance of the
reconstructed (N\'eel) phase from the physically interesting region of the
phase diagram, and, therefore, to an even larger DOF phase.
(We will further discuss the roles of $V_1^{\perp }$ and $V_1^{\parallel}$ in
stabilizing the DOF phase later in this Section). 

%++++++++++++++++++++++++++++++++++++++++++++++++++++++++++++++++++++++
\subsection{Infinite or finite roughening temperature? The role of 
$t_{ex}$.} \label{XY:sec}
%++++++++++++++++++++++++++++++++++++++++++++++++++++++++++++++++++++++

In order to discuss the point concerning the roughening temperature, 
we observe that, as $T\to \infty$, the kinetic terms tend to be the only 
relevant pieces of the Hamiltonian, see Eq.\ (\ref{parameters}). 
Thus, we now consider the model in absence of 
potential terms ($V$), and for zero chemical potential $\mu$. 
For the time being, we also take $t_{ex}=0$. 
The crucial role of the $t_{ex}$-term will be discussed afterwards.
In this case, the Hamiltonian reduces to that of two coupled XY chains:
\begin{equation} \label{XY_ham:eqn}
H=\,-t^{\parallel} \sum_{i,\alpha} 
   (S_{i,\alpha }^{+}S_{i+1,\alpha }^{-}+{\rm{H.c.}})
-t_0^{\perp} \sum_i (S_{i,1}^{+}S_{i,2}^{-}+{\rm{H.c.}})
-t_1^{\perp} \sum_{i,\alpha} (S_{i,\alpha}^+ S_{i+1,\bar{\alpha}}^- +
{\rm{H.c.}}).
\end{equation}
Notice that exchanging $t^{\parallel}$ with $t_1^{\perp}$ is simply
equivalent to renaming the sites $\left( 2i,\sigma \right)$ to 
$\left( 2i,\overline{\sigma }\right)$ and vice-versa. 
This is illustrated pictorially in Fig.\ \ref{mapp_xy:fig}.
Thus, a model with $t_1^{\perp }/t^{\parallel }=\tau $ is completely
equivalent to one with $t_1^{\perp}/t^{\parallel}=1/\tau$.

In Fig.\ \ref{gap_xy:fig}(a) we plot the finite-size gaps as a function of the
system size $L$ for $t_0^{\perp }=0$, and different values of 
$t_1^{\perp}/t^{\parallel}$ between $0$ and $1$. 
Given the negligible curvature of the straight-line fits, the data seem 
to suggest that the gap extrapolates to $0$ in all cases.
In Fig.\ \ref{k_xy:fig} we plot the finite-size value of the Luttinger 
exponent $K_S$, determined as explained in sec.\ \ref{XY:sec}.
The data for $K_S$ confirm the scenario of a gapless (i.e., rough) system. 
Notice that $K_S$ seems to extrapolate to values larger than $1$, indicating 
a rough phase that should survive to the turning on of a suitably small
repulsive $V_1^{\parallel}$.

We now address this point in more detail.
Consider the $t_1^{\perp}=t^{\parallel}$ case (with $t_0^{\perp}=0$), 
in which $K_S$ seems to extrapolate to the largest value. Denoting by 
$\left( \begin{array}{l}
         s \\ 
         s^{\prime }
        \end{array}
\right),$ with $s,s^{\prime }=\pm 1$, the four possible configurations at each
site, we can define the following four states: 
\[
\left| \uparrow \right\rangle =\left| \left( 
\begin{array}{l}
+1 \\ 
+1
\end{array}
\right) \right\rangle ,\hspace{0.3cm}\left| \downarrow \right\rangle =\left|
\left( 
\begin{array}{l}
-1 \\ 
-1
\end{array}
\right) \right\rangle ,\hspace{0.3cm}\left| 0_{\pm }\right\rangle =\frac 1{%
\sqrt{2}}\left( \left| \left( 
\begin{array}{l}
+1 \\ 
-1
\end{array}
\right) \right\rangle \pm \left| \left( 
\begin{array}{l}
-1 \\ 
+1
\end{array}
\right) \right\rangle \right) . 
\]
It is now straightforward to verify that, at each site, the state $|0_-\rangle$
is decoupled from the three remaining ones.
The Hamiltonian $H$ can then be considered a
spin-1 Hamiltonian acting on the subspace spanned by the three states 
$\left| \uparrow \right\rangle$, $\left| \downarrow \right\rangle$, and 
$\left| 0_{+}\right\rangle$. 
As it can be checked by explicitly calculating all matrix elements, {\em $H$, 
when restricted within this subspace, coincides with the Heisenberg
spin-1 Hamiltonian at $J_z=0$ and $J_{xy}=2t^{\parallel}$\/}. 
As argued by den Nijs and Rommelse,\cite{MdN_Rom} the location of the KT 
transition in the Heisenberg spin-1 phase diagram is, very likely, exactly at 
$J_z=0$.
Thus, two coupled XY chains with $t_1^{\perp}=t^{\parallel}$ and 
$t_0^{\perp}=0$ have a $K_S$ actually equal to $1$, and what we see in 
Fig.\ \ref{k_xy:fig} is only due, very likely, to finite-size effects.

The effect of turning on $t_0^{\perp}$, while keeping $t_1^{\perp}=0$, leads 
to a completely different picture. 
In this case, for $t_0^{\perp}=t^{\parallel}$ the system is gapped, 
as suggested by the finite-size data of Fig.\ \ref{gap_xy:fig}(b).
The physical reason for the different behavior of the $t_1^{\perp}$
and $t_0^{\perp}$ terms can be understood by considering the limiting cases 
of large values for these parameters. 
For $t_0^{\perp}\rightarrow \infty$, the ground state tends to 
have $\left| 0_+ \right \rangle$ at each site, with a large gap  
(of order $t_0^{\perp}$) to other exited states. 
For $t_1^{\perp}\rightarrow \infty$ (at $t_0^{\perp}=0$), on the other hand, 
the system reduces, by the previously described duality property, to two 
uncoupled XY chains, and must, therefore, be gapless.

The previous considerations lead us to conjecture that for any choice of 
$t_1^{\perp}$ (as long as $t_0^{\perp}=0$), two XY chains are gapless and
have $K_S=1$. 
On the contrary, turning on $t_0^{\perp}$, at $t_1^{\perp}=0$, immediately 
opens up a gap.
These conclusions have important consequences on the stability of the
rough phase. Since $K_S$ attains, at best, the marginal value of $1$, 
turning on any positive $V_1^{\parallel}$ immediately opens up a gap,  
and the rough phase is confined to infinite temperature.

We will now demonstrate that, if we allow for the possibility of step-crossing 
events, $t_{ex}>0$, the gapless phase survives the turning-on of a positive
$V_1^{\parallel}$, and every ``surface'' has a rough phase for high enough $T$.

In order to show this, we add to the Heisenberg spin-1 Hamiltonian a
$t_{ex}$ term, see Eqs.\ (\ref{hamiltonian},\ref{tex:eqn}), 
\[
-t_{ex}\sum_i [(S_i^{+})^2(S_{i+1}^{-})^2+\rm{H.c.}] \;, 
\]
with $t_{ex}=t^{\parallel}=J_{xy}$. 
Fig.\ \ref{h_te_ph:fig} shows the phase diagram for this case.
Qualitatively, it is very similar to the Heisenberg spin-1 case
(see Fig.\ \ref{h_ph:fig}), except for small values of the potentials, where
the $t_{ex}$ term changes the structure of the phase diagram. 
In fact, for $\mu=0$ we observe a gapless phase extending for
positive values of $V_1^{\parallel}$, up to $V_1^{\parallel}\approx 0.4$: 
this is demonstrated in Fig.\ \ref{h_te:fig} were we plot the Luttinger 
exponent $K_S$ along the line $\mu=0$.
This finding is in accord with bosonization: the $t_{ex}$ term, unlike
the $t^{\perp}$ terms, increases $K_S$ and leads to a stabilization
of the rough phase. 
Indeed, for two XY-chains it is easy to show that, up to lowest order in 
$t_{ex}$,  
\[ K_S = 1 + \frac{t_{ex}}{8\pi^2 t^{\parallel }} \;. \]

Another remarkable feature of the phase diagram in Fig.\ \ref{h_te_ph:fig}
is that, at variance with the ordinary Heisenberg case ($t_{ex}=0$), 
the temperature line for a given ``surface'' crosses the DOF
region only if the cost of a step, $\delta _S$, is sufficiently small
as compared to the interaction between steps, $\widetilde{V}_1^{\parallel}$. 
We have illustrated this by sketching in Fig.\ \ref{h_te_ph:fig} 
temperature-lines for three different situations. 
For the case labeled A, the energy cost of a step is high with respect to 
the interaction energy between steps, and there is no preroughening. 
In the case labeled B, $\delta _S/\widetilde{V}_1^{\parallel}$
is smaller, and a DOF phase is present at intermediate temperatures. 
Finally, for case C, the interaction between steps is the most relevant 
energy, and the low temperature phase is $2\times 1$ reconstructed. 

%++++++++++++++++++++++++++++++++++++++++++++++++++++++++++++++++++
\subsection{Presence or absence of preroughening: Role of interactions 
versus step line tension.}
\label{special_plane:sec}
%++++++++++++++++++++++++++++++++++++++++++++++++++++++++++++++++++

We now want to discuss in some detail what happens if we leave 
the condition $V_1^{\perp }=-V_1^{\parallel}$, without going to the
extreme case $V_1^{\perp }=0$, discussed in Sec.\ \ref{V1_perp_0:sec}.
We illustrate this by choosing $V_1^{\perp }=-V_1^{\parallel }/10$, 
while keeping $t_{ex}=t^{\parallel}=1$ and $V_0^{\perp}=\infty$ 
(infinite on-site repulsion of opposite steps).
This choice of parameters describes a class of surfaces
in which the attraction between steps of opposite kind is much smaller than
the repulsion between steps of the same kind.

In Fig.\ \ref{te_ph:fig} we plot the phase diagram for this choice of
parameters. 
The system is N\'{e}el ordered for very large values of $V_1^{\parallel}$; 
the value of the ratio $V_1^{\parallel}/|V_1^{\perp}|$ determines the location 
of the DOF-N\'{e}el phase boundary (see Fig.\ \ref{h_te_ph:fig} 
and Fig.\ \ref{pd_vpar:fig}; recall that, for $V_1^{\perp}=0$, 
the N\'{e}el phase is absent for physical values of $\mu$).
The most relevant comment to the phase diagram in Fig.\ \ref{te_ph:fig}
regards the conditions upon which the temperature trajectories of an actual
surface model cross the preroughening line. 
It is clear, in fact, that depending on the ratio between the cost of a step
(per unit length) $\delta_S$ and, say, the interaction energy between steps of
the same kind $\widetilde{V}_1^{\parallel}$, a surface can have: 
i) only roughening (case A), or ii) first preroughening and then roughening
(case B). 
Now, with $V_1^{\perp }=-V_1^{\parallel}/10$, the ``critical ratio'' 
$(\delta_S/\widetilde{V}_1^{\parallel})_{\rm crit}$, below which 
preroughening is possible is of the order of $1/10$, much smaller then in 
the $V_1^{\perp}=-V_1^{\parallel }$ case 
(where $(\delta_S/\widetilde{V}_1^{\parallel})_{\rm crit}\approx 1$).
Given the fact that $\delta_S$ is typically the largest ``diagonal'' energy, 
this implies that a physical temperature trajectory will be, most likely, 
in the region where only roughening occurs. 
If, and how, long-range interactions might change this picture is an 
interesting and open problem.

%+++++++++++++++++++++++++++++++++++++++++++++++++++++++++++++++++++
\subsection{Role of opposite-step on-site repulsion: Finite $V^{\perp}_0$} 
\label{V0_perp:sec}
%+++++++++++++++++++++++++++++++++++++++++++++++++++++++++++++++++++

At last, we want to discuss briefly what happens to the DOF phase 
if we allow double occupation of a site, i.e., if we do not take the limit
$V^{\perp}_0 \to \infty$. 
To demonstrate that the restriction to $V_0^{\perp}=\infty$ is not
crucial, we consider the case $V^{\parallel}_1=-V^{\perp}_1=1$,
$t^{\parallel}=t^{\perp}_1=1$, with $V^{\perp}_0$ finite. 
If $V^{\perp}_0 \to \infty$ the system is an Heisenberg spin-1 chain at the 
isotropic point, corresponding to a DOF phase.\cite{MdN_Rom}
In Fig.\ \ref{to_heis:fig} we plot the finite-size values of the flatness
order parameter ${\cal F}$ (open symbols) and of the DOF correlation
function $G_s(L/2)$ (full symbols) for decreasing values of $V^{\perp}_0$. 
The data suggest that the system remains DOF all the way down to 
$V^{\perp}_0 \approx 0$. 
We have verified that a similar scenario is found if we turn on the
$t_{ex}$ term or a small $|\mu|$.
Thus, our finite-size data suggest that the spin-1 condition 
($V_0^{\perp}=\infty$) is not essential in order to stabilize a DOF phase. 

%+++++++++++++++++++++++++++++++++++++++++++++++++++++++++++++++++++
\section{Step-step correlations} \label{corr:sec}
%+++++++++++++++++++++++++++++++++++++++++++++++++++++++++++++++++++

Correlation functions involving steps can be calculated numerically, 
for a given finite size, at any point in the phase diagram of our model. 
We will discuss here two correlation functions, i.e., 
step-step correlations and terrace width distributions. 
Let $n_S$ be the average density of steps of a single species (up or down). 
In general $n_S$ is always different from zero, even in the flat phase, since
we do not discriminate between steps that traverse the entire sample and
steps that form loops (i.e., finite terraces). 
Step-step correlations are defined as follows:
\begin{equation}
N^{\uparrow \sigma}(r) = \frac{1}{n_S^2} 
    \langle {\rm Step}^{\uparrow}(0) {\rm Step}^{\sigma}(r) \rangle = 
\frac{1}{n_S^2} \langle n_{0,\uparrow} n_{r,\sigma} \rangle  \;,
\end{equation}
with $\sigma=\uparrow,\downarrow$. 
If translational symmetry is not broken, we must have, at large distances, 
$N^{\uparrow \sigma}(r\to \infty ) \to 1$. 
The distribution of terrace sizes (along the $x$-direction only!) is the 
probability of having two steps a distance $r$ apart without any other 
step in between. 
There are two different kind of terraces we can look at: 
those delimited by two steps of the same type, and those between two different 
steps. Thus, we define:
\begin{equation}
P^{\uparrow \sigma}(r) = \frac{1}{n_S^2} \langle n_{0,\uparrow} 
\left[ \prod_{j=1}^{r-1} (1-n_{j,\uparrow})(1-n_{j,\downarrow}) \right]
n_{r,\sigma} \rangle \;, 
\end{equation}
where, again, $\sigma=\uparrow,\downarrow$. 
The string operator in square brackets enforces the absence of additional steps 
between $0$ and $r$. 
Fig.\ \ref{corr_ss:fig} illustrates the behavior of 
$N^{\uparrow \uparrow}$ and $N^{\uparrow \downarrow}$, 
at three different points in the phase diagram of the Heisenberg spin-1 chain: 
a rough case ($J_z=-0.5$, $\mu=0$, triangles), 
a DOF case ($J_z=1$ and $\mu=0$, squares),
and a flat one ($J_z=1$ and $\mu=-2$, pentagons). 
The flat case results are very simple: 
both $N^{\uparrow \uparrow}$ and $N^{\uparrow \downarrow}$
converge exponentially fast (with a very short correlation length) to
the large distance limit of $1$. 
In the rough phase, instead, we have verified that the approach to $1$ 
shows a power law tail. This is easy to prove. 
Rewrite first $N^{\uparrow \sigma}$ in terms of density and ``spin'' 
correlations:
\begin{equation}
N^{\uparrow \sigma}(r) = \frac{1}{4n_S^2} \left[
    \langle n_0 n_r \rangle  \pm 
    \langle S^{z}_0 S^{z}_r \rangle  \right] \;,
\end{equation}
where the $+$ and $-$ signs apply, respectively, to $\sigma=\uparrow$ and
$\sigma=\downarrow$, $n_i=n_{i,\uparrow}+n_{i,\downarrow}$, and 
$S^z_i=n_{i,\uparrow}-n_{i,\downarrow}$.
Within a bosonization approach,\cite{Strong} the operators $n_i$ and $S^z_i$ 
involve (after particle-hole transformation for the $\downarrow$-bosons) only 
the antisymmetric and symmetric sectors, respectively. 
The antisymmetric sector is always gapped (see discussion in Sec.\ \ref{K:sec} 
and Ref.\ \cite{Strong}), so that density-density correlations are exponential. 
In the rough phase, however, the symmetric sector is gapless, and 
$S^z-S^z$ correlations have a uniform power-law tail of the form
\begin{equation}
\langle S^{z}_0 S^{z}_r \rangle \,=\, -\frac{K_S}{(\pi r)^2} \,+\, \cdots \;,
\end{equation}
which is precisely the term responsible for the logarithm in the heigth-heigth
correlation function $G(r)=\langle(h_r-h_0)^2\rangle$.\cite{G_h:note}
The DOF case results, finally, show a different behavior, with a sizeable 
oscillating component of the correlations. 
This behavior, however, reflects only a short-range effect, caused by
the neighboring reconstructed (N\'eel) phase: 
the oscillating part has to decrease to zero at large $r$, since no breaking of 
translational symmetry occurs in the DOF phase.\cite{Rom_MdN}

We finally discuss briefly the behavior of the distributions of terrace
sizes (for simplicity, once again, in the Heisenberg spin-1 case).
While in principle it is important to know what is the probability for
the surface to be flat over a distance $r$, this quantity has never been
calculated so far. 

Let us consider, first, the behavior of $P^{\uparrow \downarrow}$ in the
rough phase. Fig.\ \ref{corr_terr:fig}(a) is a plot of the logarithm of 
$P^{\uparrow \downarrow}$ versus a scaled distance $2n_Sr$,
for several points taken inside the rough phase of the Heisenberg spin-1 
phase diagram. 
We observe that the general behavior of $P(r)$ is exponential in the
size of the terrace $r$, 
\[ P(r) \approx e^{-r/\lambda} \]
and that a good collapse is obtained for all data
if the distance $r$ is scaled to the average separation between two steps, 
$1/(2n_S)$, i.e., $\lambda\propto 1/(2n_S)$.  
The scattering of the data for the largest $r$'s is due to
finite-size effects. 
The behavior of $P^{\uparrow \uparrow}(r)$ is found to be qualitatively 
similar.

In the DOF phase we find that the terrace size distribution probability is again
exponential with size, but now $\lambda$ does not scale with the density
of steps, as it did instead in the rough phase. 
Fig.\ \ref{corr_terr:fig}(b) illustrates the behavior of $P(r)$ at a DOF point,
corresponding to the isotropic Heisenberg point of the spin-1 chain. 
$P(r)$ at a rough point is also reported for comparison. 
We observe that, as anticipated, the behavior of $P^{\uparrow \sigma}$ is, once
again, exponential in $r$ 
(i.e., the ``DOF checkerboard'' has no typical length!).
Superimposed on the leading exponential, the DOF case results show a strong 
oscillating short range component which is again due to the neighboring 
reconstructed (N\'eel) phase. 
Two more features are worth noticing. First, compared to the rough case, 
$P^{\uparrow \downarrow}(r)$ is larger in the DOF case for $r=1$, and then 
substantially smaller for larger values of $r$ (and decreasing with a 
larger exponent). Second, in the DOF case $P^{\uparrow \uparrow}(r)$ is one 
order of magnitude smaller than $P^{\uparrow \downarrow}(r)$, while the
difference is much smaller in the rough case. 
These features are reasonable in view of the diluted antiferromagnetic 
ordering of steps, typical of the DOF phase. 

Experimentally, terrace sizze distributions could in the future be extracted,
e.g., from STM data.\cite{hoogeman}

%+++++++++++++++++++++++++++++++++++++++++++++++++++++++++++++++++++
\section{Summary and conclusions} \label{conclusions:sec}
%+++++++++++++++++++++++++++++++++++++++++++++++++++++++++++++++++++

In this paper we have presented and discussed a statistical mechanics 
model for studying the possible phase transitions of an ideal, 
unreconstructed surface. 
The elementary objects upon which the model is based are the natural 
extended defects of an unreconstructed surface, i.e., steps and terraces.
This starting point is, in our opinion, physically more transparent then
the usual microscopic RSOS-model description.
Our model allows, in principle, the description of a real surface 
and, in perspective, one could test it with realistic step-step interactions.

We have tackled our problem of interacting steps by mapping it, in a
well known way, onto a one dimensional quantum problem of interacting
hard-core bosons. 
Although this mapping is exact only in the strong anisotropy limit, it can 
provide very useful information about the phases and the nature of the 
transitions also in more general instances. 
Moreover, some realistic cases, like (110) surfaces of fcc metals,  
are actually quite anisotropic. 

The quantum Hamiltonian, see Eq.\ (\ref{hamiltonian}), contains standard 
terms, like nearest-neighbor hopping (describing kinks on the steps), 
potential terms (describing interactions between steps), and chemical potential 
(cost per unit length of a step), as well as terms describing 
i) terrace creation/annihilation (through BCS-like number non conserving
terms), ii) opposite step crossing events.
The latter two terms are crucial, in many ways. 

Terrace terms are important to describe correctly the universality classes
of the relevant transitions. 
This is known in the literature,\cite{Bohr,Bal_Kar,MdN_92} but
never explored in details in the present context.
Moreover, in our case, the terrace terms also {\it force\/}
us to work with hard-core bosons, as the standard Wigner-Jordan transformation
to fermions does not lead to a simple local fermionic Hamiltonian. 
This point is sometimes overlooked in the literature.\cite{Kar_Sha}

The term describing the crossing of opposite steps is important in order
to stabilize a gapless (i.e., rough) phase for finite repulsive interactions 
between steps of the same kind. This, in turn, leads to a finite roughening
temperature for the classical model. 

Finite-size exact diagonalizations and bosonization techniques have been
used to unveil the richness of the phase diagram. 
In the limit of $V_0^{\perp}\to\infty$ and for a particular choice of 
parameters (the potentials, for instance, are truncated to first neighbors
and set to $V_1^{\parallel }=-V_1^{\perp }$), the model maps exactly onto
the Heisenberg spin-1 chain Hamiltonian.
The latter was also obtained, by den Nijs and Rommelse 
as the quantum mapping of RSOS models;\cite{MdN_Rom} 
it presents a DOF phase for $V_1^{\parallel}>0$, but does not 
describe, in that case, a surface with a finite temperature roughening. 
(On the other hand, if $V_1^{\parallel}$ is attractive there is only 
roughening).

Taking the Heisenberg chain as a starting point, we have then explored 
the phase diagram for other choices of parameters,
obtaining results that we believe to be relevant with respect to 
the surface physics interpretation of our model. 
Summarizing, we have seen that:

\begin{enumerate}

\item The Heisenberg spin-1 restriction $V_1^{\parallel}=-V_1^{\perp}$ 
is not crucial in stabilizing the DOF phase. In particular,
we observe a DOF phase even for $V_1^{\perp}=0$ 
(see Sec.\ \ref{V1_perp_0:sec}). 
Moreover, a DOF phase is present not only for $V_0^{\perp}=\infty$ 
(spin-1 case) but also when $V_0^{\perp}$ is finite, as long as positive. 

\item If we add to the Heisenberg spin-1 Hamiltonian a $t_{ex}$ term,
we observe a gapless phase extending for positive values of $V_1^{\parallel}$.
Every surface has a rough phase for high enough $T$. 
This is true also for other choices of the
potentials (see Sec.\ \ref{special_plane:sec}). 
Moreover, if we do not include in the Hamiltonian the $t_{ex}$ term, the
rough phase does not survive when one turns on a $V_1^{\parallel}>0$ 
(see sec.\ \ref{XY:sec}). 
Thus, the opposite step crossing term is {\it crucial\/} in order
to obtain a model describing, at least at a coarse grained level, a
physical surface.

\item 
The relative values of the interactions and of the cost per unit length 
of a step decide whether a surface has a stable DOF phase for a certain 
range of $T$.
The temperature trajectory crosses the DOF region only if the cost of a 
step, $\delta_S$, is sufficiently small as compared to 
$\widetilde{V}_1^{\parallel}$ (see Fig.\ \ref{h_te_ph:fig}). 
Given the fact that $\delta_S$ is typically the largest ``diagonal'' energy, 
this implies that a physical temperature trajectory will often be 
in a region where only roughening occurs. 

\end{enumerate}

In conclusion, we have found that: 
(i) a model based on steps can describe preroughening (PR), as well as 
roughening; 
(ii) the steps must be treated as hard-core bosons rather than fermions;
(iii) the qualitative role of step-step interactions in driving PR, 
known already from RSOS models, is recovered in this picture; 
(iv) correlation functions involving steps can be calculated in a 
quite straightforward way. 
The main one, never studied so far, which we have considered, is the terrace 
size distribution. Here we find simply an exponentially decreasing probability
for increasing size. This result should be amenable to experimental testing,
for example by STM; 
(v) In view of the additional simplicity of step models, it should be 
feasible, in the future, to study the role of long-ranged interactions, 
a problem without hope of solution within RSOS models.

ACKNOWLEDGMENTS - We thank M. Fabrizio, A. Parola, S. Sorella, and 
F.D.M. Haldane for many useful discussions. 
We acknowledge financial support from INFM, through Projects LOTUS
and HTSC, from EU, through ERBCHRXCT940438, and from MURST, through COFIN97.

%%%%%%%%%%%%%%%%%%%%%%%%%%%%%%%%%%%%%%%%%%%%%%%%%%%%%%%%%%%%%%%%%%%%%%%%
% BIBLIOGRAPHY
%%%%%%%%%%%%%%%%%%%%%%%%%%%%%%%%%%%%%%%%%%%%%%%%%%%%%%%%%%%%%%%%%%%%%%%%

%%%%%%%%%%%%%%%%%%%%%%%%%%%%%%%%%%%%%%%%%%%%%%%%%%%%%%%%%%%%%%%%%%%%%%%%
% FIGURE CAPTIONS
%%%%%%%%%%%%%%%%%%%%%%%%%%%%%%%%%%%%%%%%%%%%%%%%%%%%%%%%%%%%%%%%%%%%%%%%
\begin{center}
{\bf FIGURE CAPTIONS}
\end{center}

\begin{figure}
\caption{
Scheme of a surface with up ($\uparrow$) and down ($\downarrow$) steps. 
The heights of the terraces are explicitly indicated. 
The other symbols refer to the Boltzmann weights considered: 
$\delta_K$ (cost of a kink), $\delta_T$ (cost of a terrace creation), 
$\delta_{ex}$ (cost of a step crossing), 
$V^{\parallel}$ and $V^{\perp}$ (interactions between parallel and opposite
steps); the black dots indicate where terraces are created or destroyed.}
\label{step:fig}
\end{figure}

\begin{figure}
\caption{
Schematic representation of a kink, the beginning of a size-1 terrace, 
a size-0 terrace,
and a step crossing between strip $j$ and strip $j+1$ and the relative 
energetic costs.}
\label{transfer:fig} 
\end{figure}

\begin{figure}
\caption{
Phase diagram for the spin-1 Heisenberg chain.\protect\cite{MdN_Rom} 
Here and in the following we consider only negative values of the chemical 
potential $\mu$, that are relevant for the surface physics problem. 
Lines (a) and (b) are discussed in the text.}
\label{h_ph:fig} 
\end{figure}

\begin{figure}
\caption{Qualitative phase diagram for a spin-1 chain with 
$t_1^{\perp}=t^{\parallel}=1$, $t_{ex}=0$, $V_1^{\perp}=0$.}
\label{pd_vpar:fig} 
\end{figure}

\begin{figure}
\caption{
The duality mapping for two XY chains, see Eq.\ (\protect\ref{XY_ham:eqn}). 
Dotted, solid and dashed lines denote, respectively, $t_1^{\perp}$, 
$t^{\parallel}$, and $t_0^{\perp}$ couplings. 
The duality survives also in presence of $V_0^{\perp}$.}
\label{mapp_xy:fig} 
\end{figure}

\begin{figure}
\caption{
(a) Finite-size charge excitation gaps for the two XY chains 
in Eq.\ (\protect\ref{XY_ham:eqn}) at $t_0^{\perp }=0$, 
$t^{\parallel}=1$, and for various values of $t_1^{\perp}$. 
Dashed lines are obtained from straight lines constructed so as to
pass through the $L=6$ and $L=8$ points. 
The extrapolation to zero is remarkably good.
(b) Finite-size charge excitation gaps for $t_0^{\perp}=0$, 
$t_1^{\perp}=t^{\parallel}=1$ (dashed line), and $t_1^{\perp }=0$, 
$t_0^{\perp}=t^{\parallel}=1$ (solid line).
The solid and dashed straight lines are constructed as in (a). 
Notice the remarkable smallness of size corrections. }
\label{gap_xy:fig} 
\end{figure}

\begin{figure}
\caption{Luttinger exponent $K_S$ for the same parameters of 
Fig.\ \protect\ref{gap_xy:fig}(a). 
As argued in the text (see sec.\ \protect\ref{XY:sec}), $K_S$ should converge to
$1$ for $L\to \infty$ and the apparent extrapolation
to values larger than $1$ is very likely due to finite size effects.}
\label{k_xy:fig} 
\end{figure}

\begin{figure}
\caption{Qualitative phase diagram for a Heisenberg spin-1 chain with 
exchange term, i.e., $t_1^{\perp }=t^{\parallel }=t_{ex}=1$, 
$V_1^{\parallel}=-V_1^{\perp}$. Lines (A), (B), and (C) are 
discussed in the text.}
\label{h_te_ph:fig} 
\end{figure}

\begin{figure}
\caption{Finite-size Luttinger exponent $K_S$ along the line $\mu=0$, for
the Heisenberg spin-1 chain plus exchange term. $K_S$ extrapolates to values
larger than $1$ for $V_1^{\parallel }<0.4$. 
Dashed lines are only guides to the eye.}
\label{h_te:fig} 
\end{figure}

\begin{figure}
\caption{Qualitative phase diagram for a spin-1 chain with 
$t_1^{\perp}=t^{\parallel}=t_{ex}=1$, $V_1^{\perp}=-V_1^{\parallel}/10$.
Line (A) describes a situation where the cost of a step $\delta_S$ is larger
than the interactions, and only roughening is found. Line (B) describes
a situation where $\delta_S$ is smaller than $|\widetilde{V}_1^{\perp}|$, and
roughening is preceded by PR.}
\label{te_ph:fig} 
\end{figure}

\begin{figure}
\caption{Finite-size values of the flatness order parameter ${\cal F}$ 
(open symbols) and of the DOF correlation function $G_s(L/2)$ (full symbols) 
at the Heisenberg isotropic point, for decreasing values of $V^{\perp}_0$.
The system appears to be DOF for all positive values of $V^{\perp}_0$.
The dashed lines are only guides to the eye.}
\label{to_heis:fig} 
\end{figure}

\begin{figure}
\caption{
$N^{\uparrow \downarrow}$ (above) and $N^{\uparrow \uparrow}$ (below), 
at three different points in the phase diagram of the Heisenberg spin-1 chain: 
a rough case ($J_z=V^{\perp}=-V^{\parallel}_1=-0.5$, $D=-\mu=0$, triangles), 
a DOF case ($J_z=V^{\perp}=-V^{\parallel}_1=1$, and $D=-\mu=0$, squares),
and a flat one ($J_z=V^{\perp}=-V^{\parallel}_1=1$, and $D=-\mu=2$, pentagons).
Lines are only guides to the eye. } 
\label{corr_ss:fig}
\end{figure}
 
\begin{figure}
\caption{
(a) $ln(P^{\uparrow \downarrow})$ versus the scaled distance $2n_Sr$
at various points in the rough phase of the Heisenberg spin-1 chain; 
full symbols: $J_z=V^{\perp}=-V^{\parallel}_1=-0.25$, $D=-\mu=0,0.1,0.2,0.3$;
empty symbols: $J_z=V^{\perp}=-V^{\parallel}_1=-0.5$, $D=-\mu=0,0.2,0.4$;
stars: $J_z=V^{\perp}=-V^{\parallel}_1=-0.75$, $D=-\mu=0,0.1,0.2,0.3$.
(b) $P^{\uparrow \downarrow}$ and $P^{\uparrow \uparrow }$, in logarithmic
scale, at two different points in the phase diagram of the Heisenberg spin-1 
chain: a DOF case ($J_z=V^{\perp}=-V^{\parallel}_1=1$, and $D=-\mu=0$, squares),
and a rough case ($J_z=V^{\perp}=-V^{\parallel}_1=-0.5$, $D=-\mu=0$, 
triangles). Full and empty symbols correspond to 
$P^{\uparrow \downarrow}$ and $P^{\uparrow \uparrow}$, respectively. }
\label{corr_terr:fig}
\end{figure}


\begin{thebibliography}{99}
\bibitem{vicinal:note}
For a vicinal surface of angle $\phi$, we would have 
$\sum_i\left( S_{i,1}^z+S_{i,2}^z\right) =L\tan \phi$.
\bibitem{G_h:note}
With the notation ${S_i^z=n}_{i\uparrow }-n_{i\downarrow },$ the height
difference between site $r$ and site $0$ is given by the sum 
$ \sum_{i=1}^{r-1}S_i^z$. Thus, using translational invariance for the
spin-spin correlation function, 
\[
G_h\left( r\right) =\sum\limits_{i,j=1}^{r-1}\left\langle
S_i^zS_j^z\right\rangle =r\left\langle \left( S_0^z\right) ^2\right\rangle
+2\sum\limits_{j=1}^{r-1}\left( r-j\right) \left\langle
S_0^zS_j^z\right\rangle \approx -2\sum\limits_{j=1}^{r-1}j\left\langle
S_0^zS_j^z\right\rangle 
\]
If the uniform part of the spin-spin correlation function 
$\langle S_0^zS_j^z\rangle$ behaves, for large $j$'s, 
like $\frac{-K}{4\pi^2j^2}$, 
$G_h(r)$ will diverge logarithmically as $\frac{2K}{\pi^2}\ln (r)$.
\end{thebibliography}
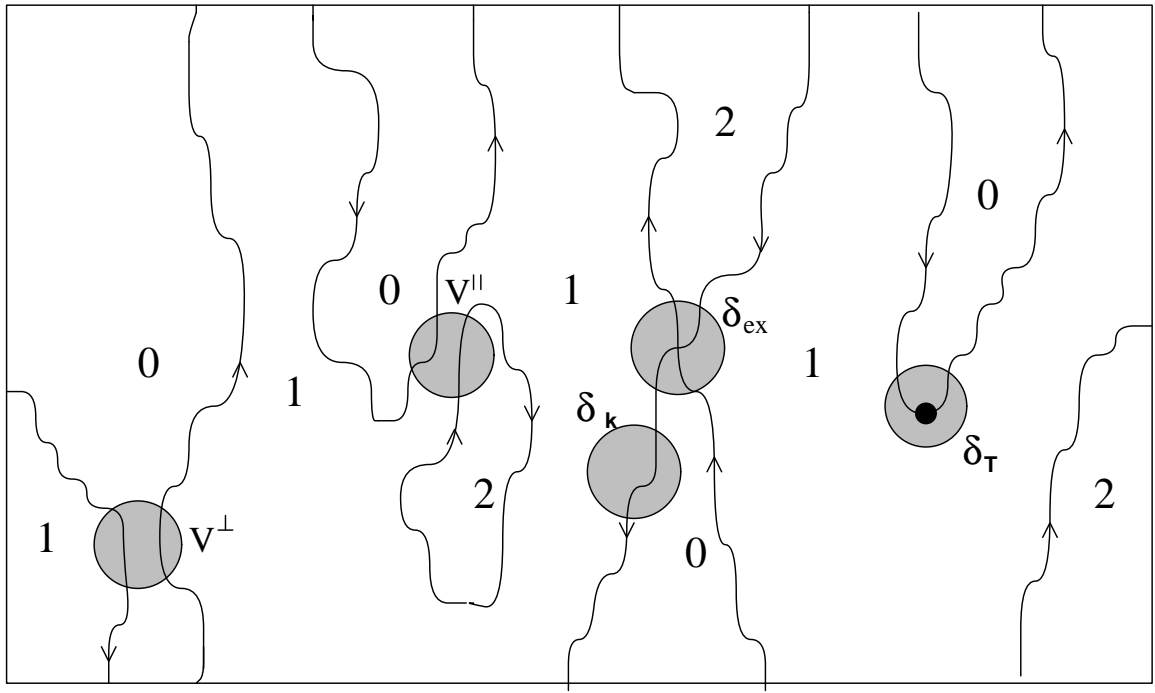
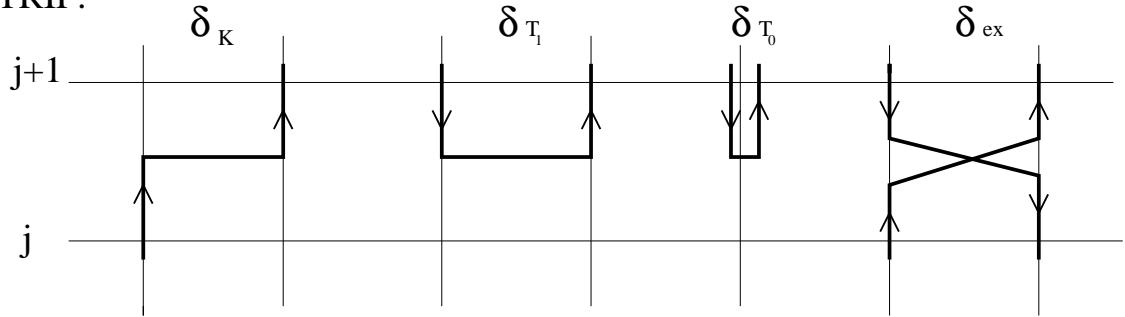
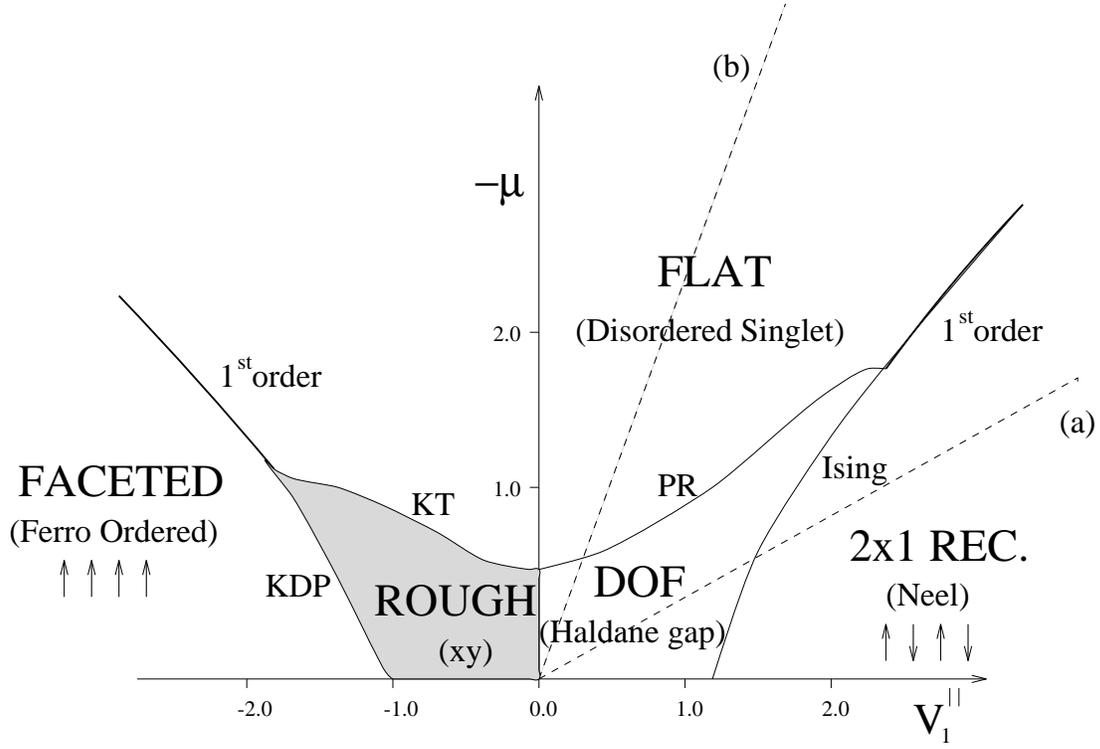
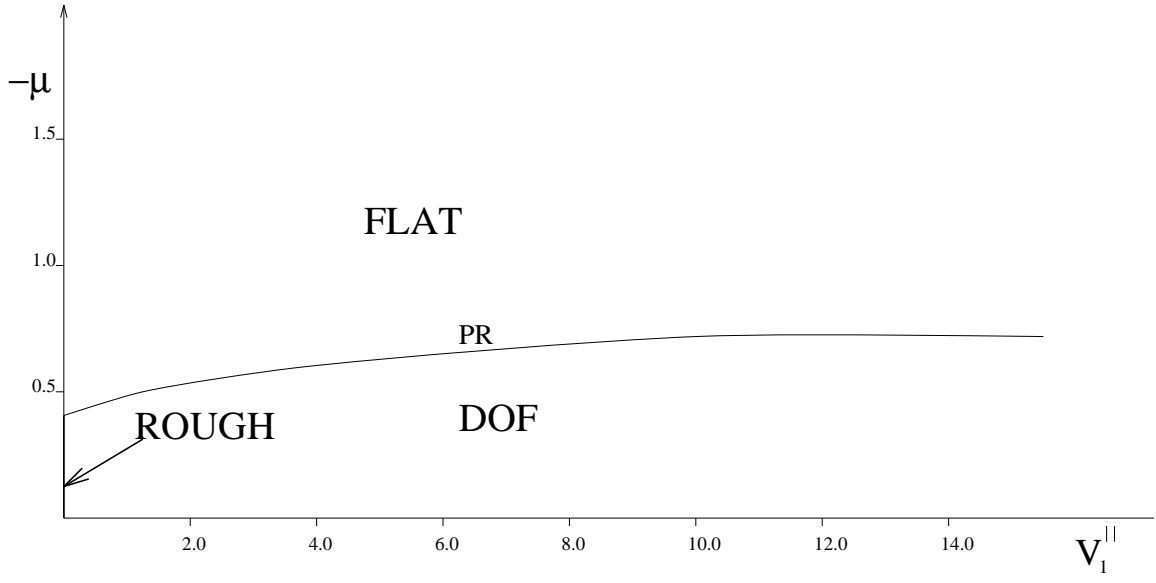
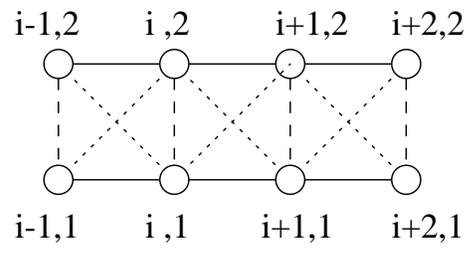
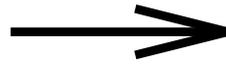
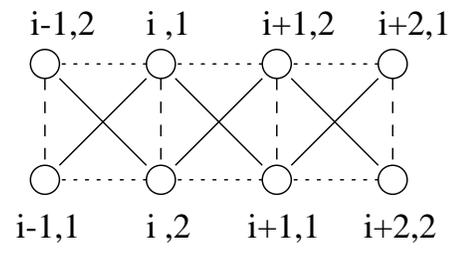
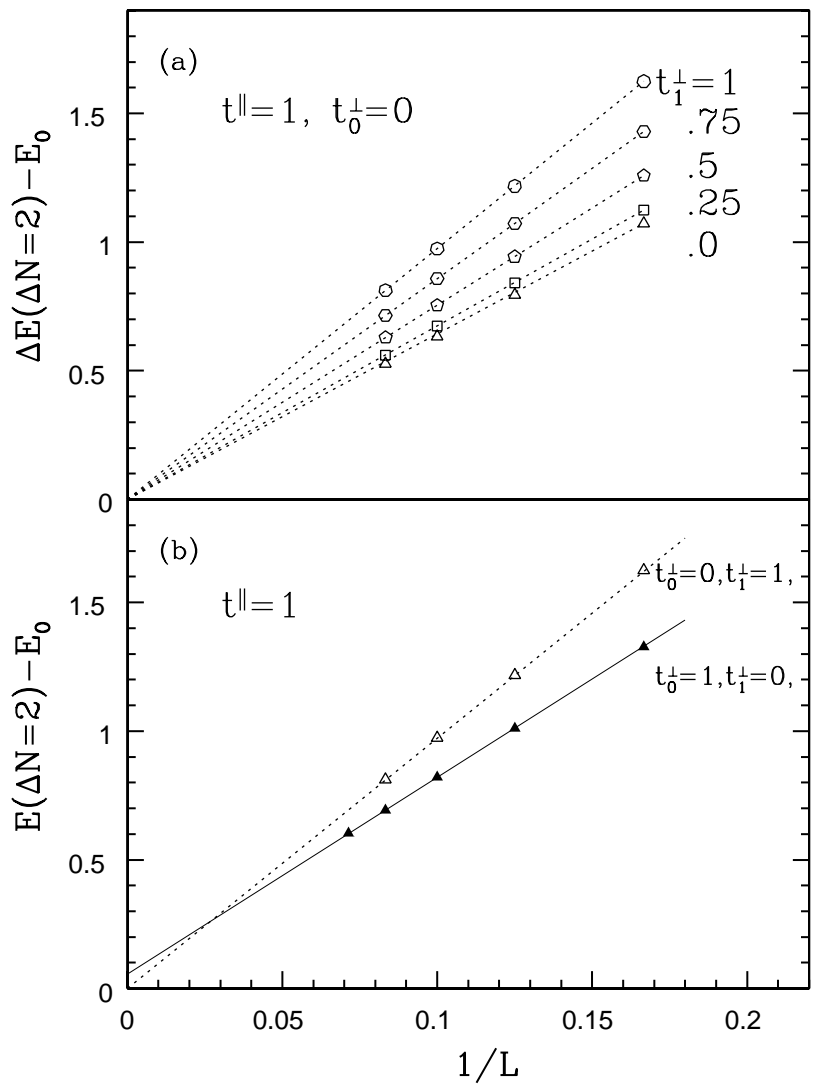
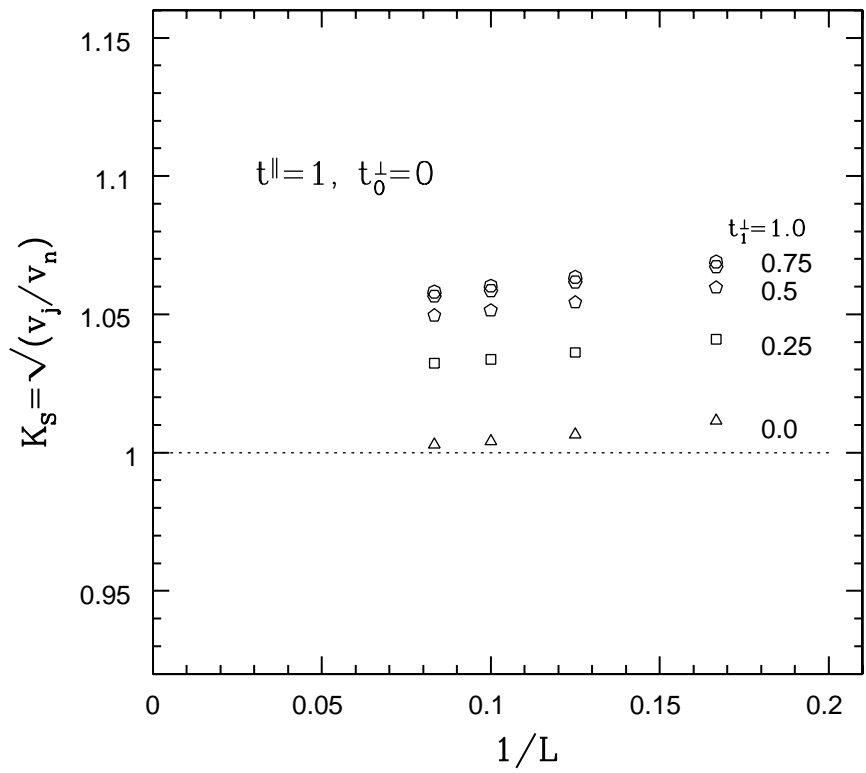
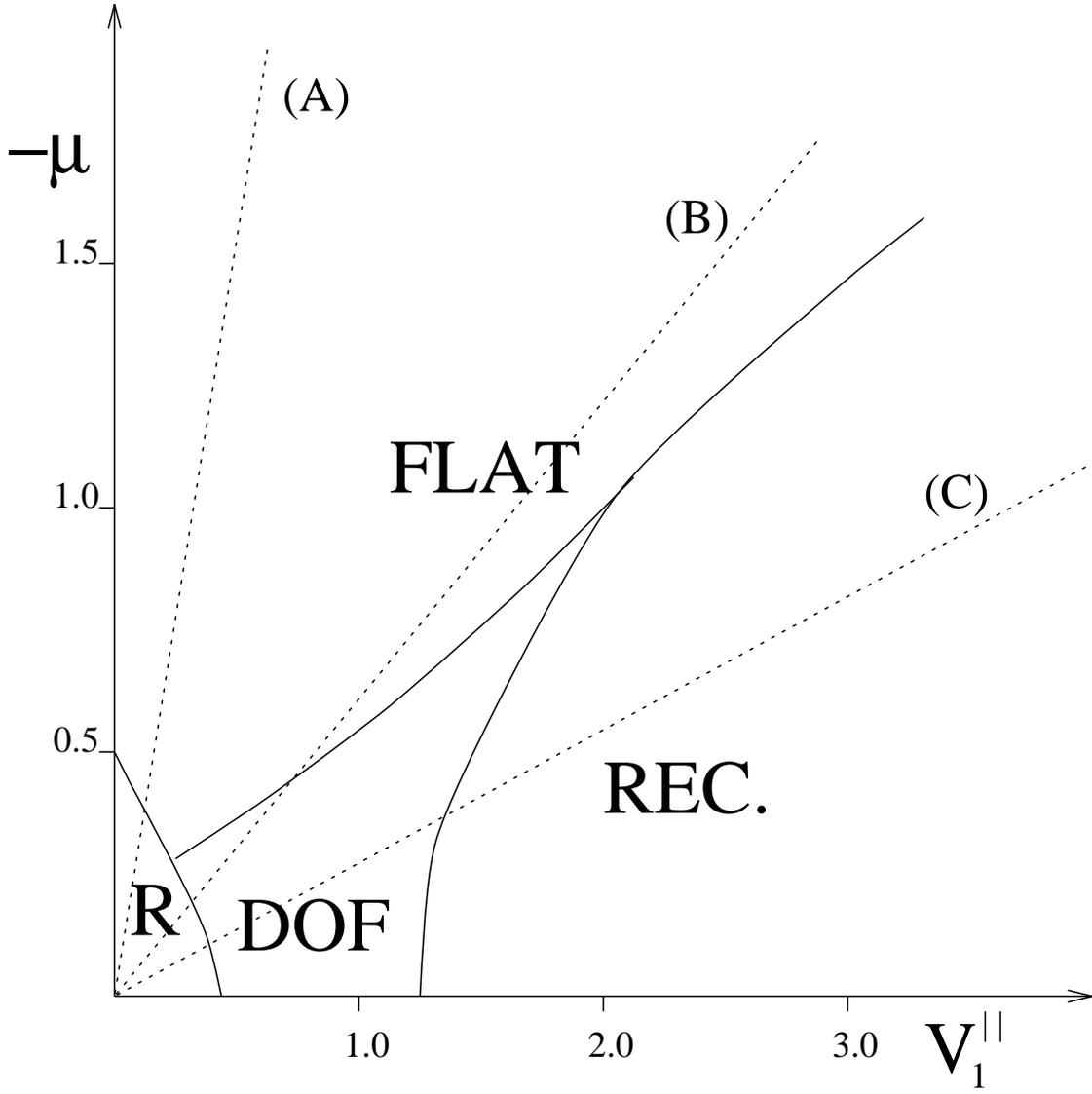
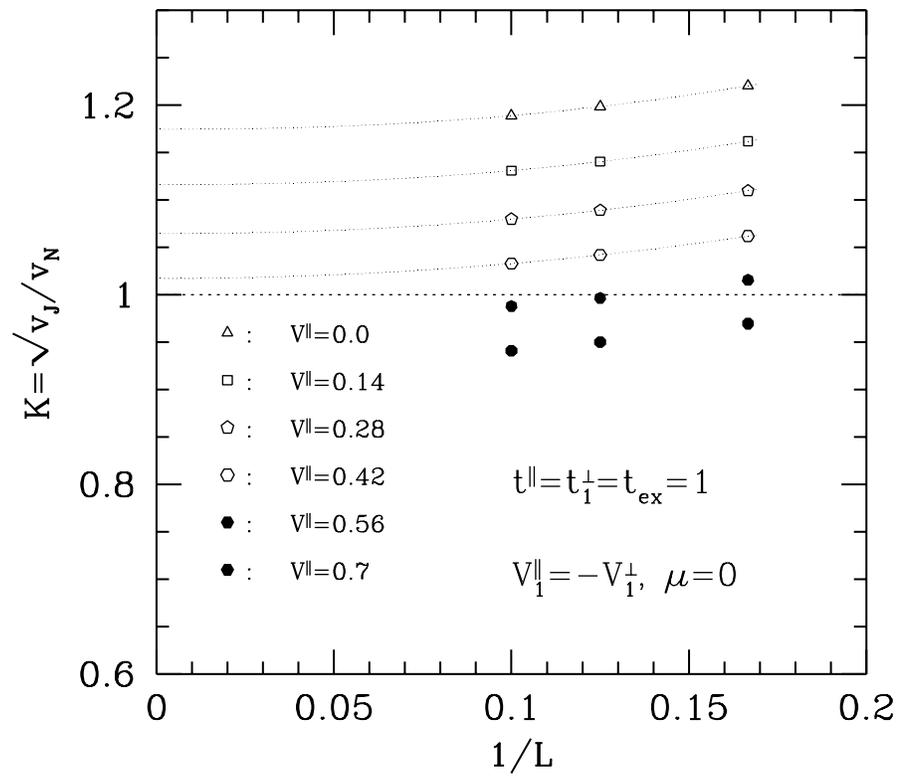
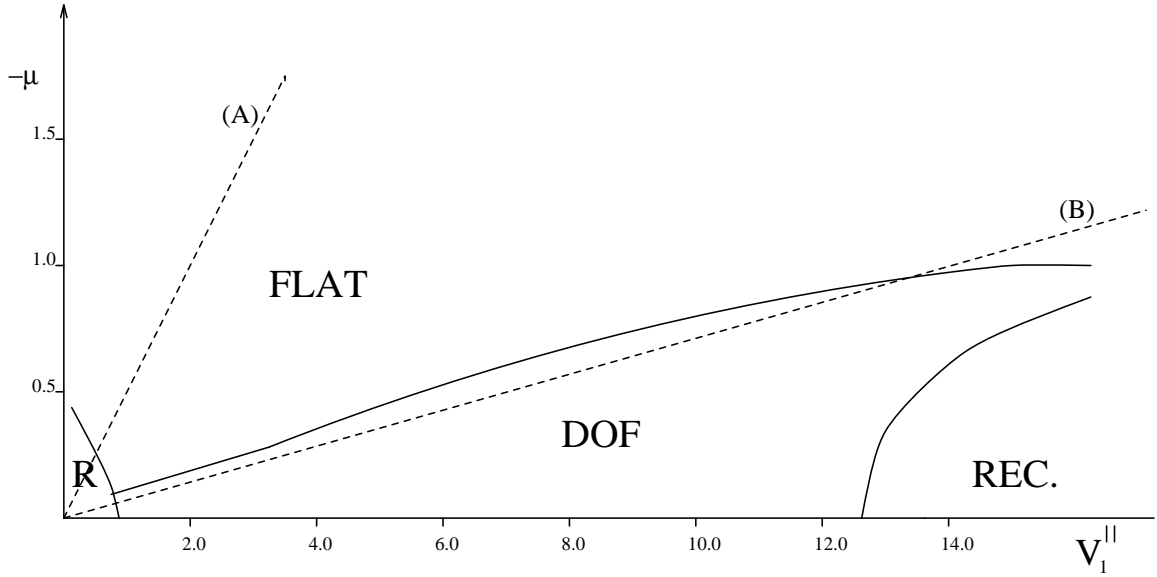
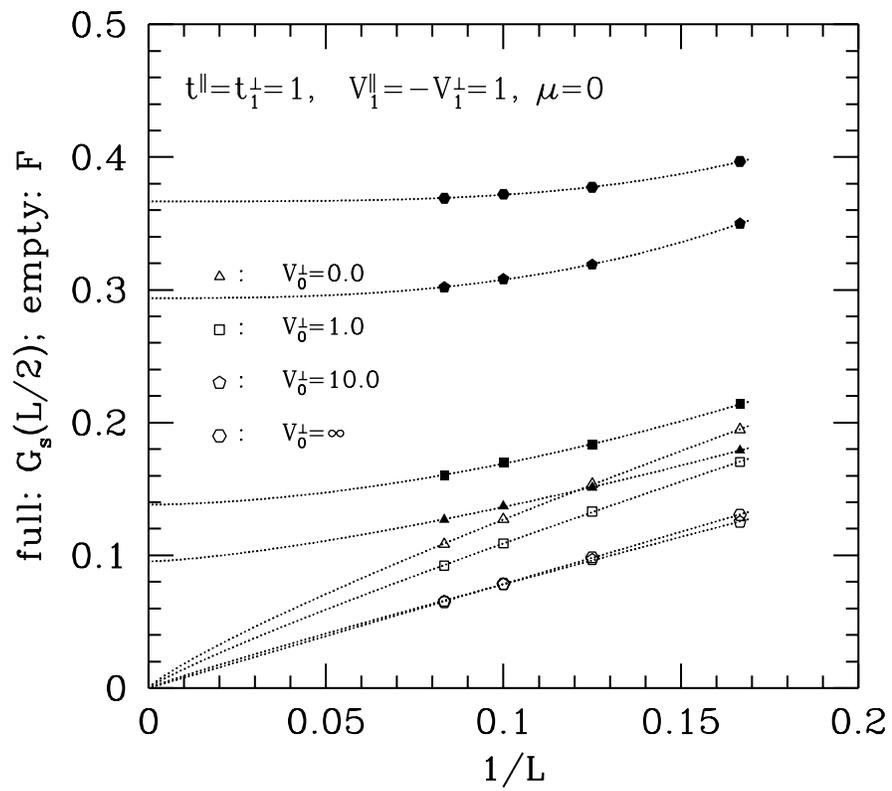
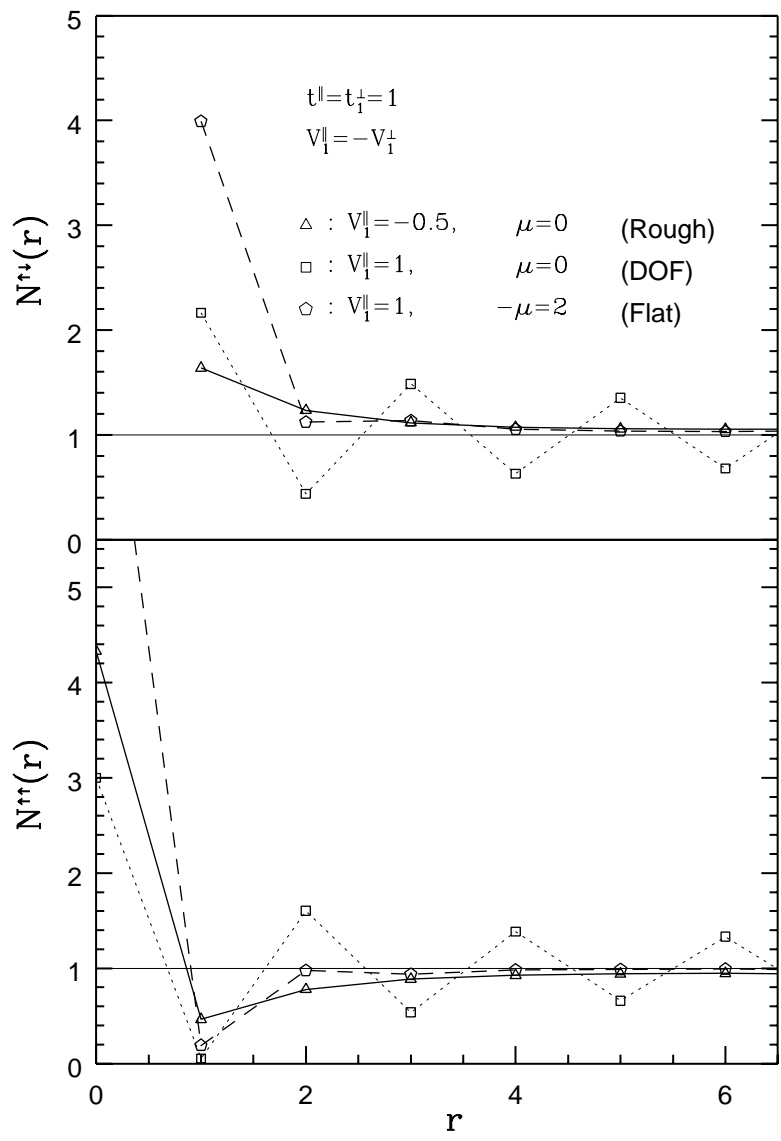
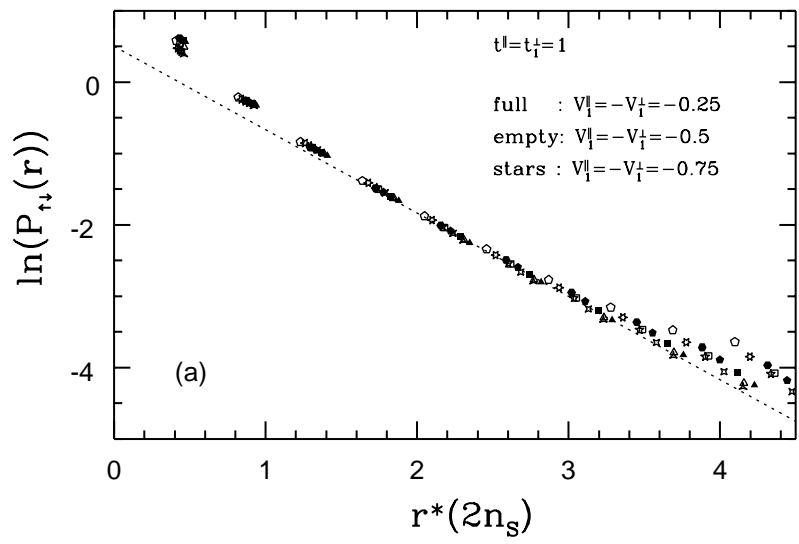
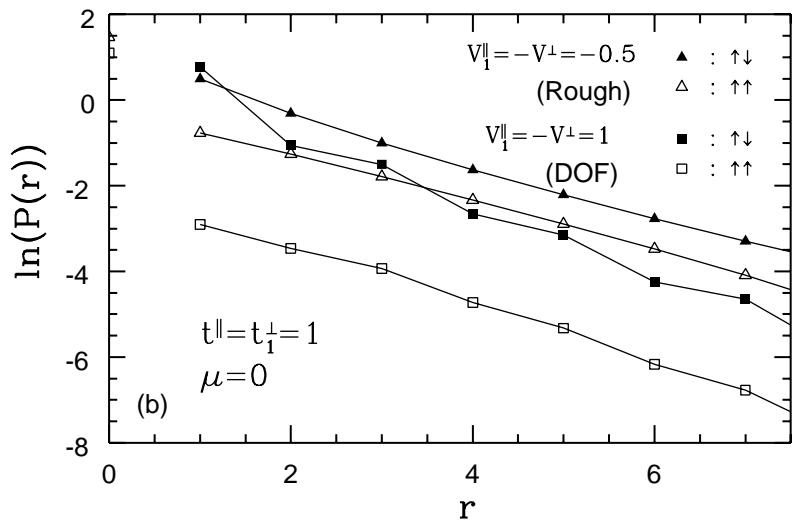
\end{document}